\documentclass[useAMS,usenatbib]{mn2e}

\usepackage{graphicx}
\usepackage{supertabular}
\usepackage{url}
\usepackage[totalwidth=530pt, totalheight=730pt]{geometry}

\newcommand{\sersic}{{S\'ersic}}
\newcommand{\gala}{{\sc Galapagos}}
\newcommand{\galfit}{{\sc Galfit}}
\newcommand{\gimtwod}{{\sc Gim2d}}
\newcommand{\budda}{{\sc Budda}}
\newcommand{\iraf}{{\sc Iraf}}
\newcommand{\sex}{{\sc SExtractor}}
\newcommand{\hst}{{\em HST}}

\newcommand{\ie}{i.e.}
\newcommand{\eg}{e.g.}

\title[GALAPAGOS: From Pixels to Parameters]
{GALAPAGOS: From Pixels to Parameters}
\author[M. Barden et al.]{Marco Barden,$^{1,2}$\thanks{E-mail:
marco.barden@uibk.ac.at (MB)} Boris H\"au\ss ler,$^{2,3}$ Chien Y.
Peng$^{4,5}$, Daniel H. McIntosh$^{6,7}$\newauthor
and Yicheng Guo$^{7}$\\
$^{1}$Institute of Astro- and Particle Physics, University of Innsbruck,
Technikerstra\ss e 25, A-6020 Innsbruck, Austria\\
$^{2}$Max-Planck-Institute for Astronomy, K\"onigstuhl 17, D-69117 Heidelberg,
Germany\\
$^{3}$Schools of Physics \& Astronomy, University of Nottingham, University
Park, Nottingham NG7 2RD, UK\\
$^{4}$NRC Herzberg Institute of Astrophysics, 5071 West Saanich Road, Victoria,
British Columbia V9E 2E7, Canada\\
$^{5}$Space Telescope Science Institute, 3700 San Martin Drive, Baltimore,
MD 21218, USA\\
$^{6}$Department of Physics, University of Missouri-Kansas City, 5100 Rockhill
Road, Kansas City, MO 64110, USA\\
$^{7}$Department of Astronomy, University of Massachusetts, 710 North Pleasant
Street, Amherst, MA 01003, USA}

\begin{document}

\date{Accepted 2012 January 23. Received 2011 December 07; in original
form 2011 July 29}

\pagerange{\pageref{firstpage}--\pageref{lastpage}} \pubyear{2012}

\maketitle

\label{firstpage}

\begin{abstract}
To automate source detection, two-dimensional light-profile \sersic\ modelling
and catalogue compilation in large survey applications, we introduce a new code
\gala, Galaxy Analysis over Large Areas: Parameter Assessment by GALFITting
Objects from SExtractor. Based on a single setup, \gala\ can process a complete
set of survey images. It detects sources in the data, estimates a local sky
background, cuts postage stamp images for all sources, prepares object masks,
performs \sersic\ fitting including neighbours and compiles all objects in a
final output catalogue. For the initial source detection \gala\ applies \sex,
while \galfit\ is incorporated for modelling \sersic\ profiles. It measures
the background sky involved in the \sersic\ fitting by means of a flux growth
curve. \gala\ determines postage stamp sizes based on \sex\ shape parameters. In
order to obtain precise model parameters \gala\ incorporates a complex sorting
mechanism and makes use of modern CPU's multiplexing capabilities. It combines
\sex\ and \galfit\ data in a single output table. When incorporating information
from overlapping tiles, \gala\ automatically removes multiple entries from
identical sources. \gala\ is programmed in the Interactive Data Language, IDL.
We test the stability and the ability to properly recover structural parameters
extensively with artificial image simulations. Moreover, we apply \gala\
successfully to the STAGES data set. For one-orbit \hst\ data, a single 2.2 GHz
CPU processes about 1000 primary sources per 24 hours. Note that \gala\ results
depend critically on the user-defined parameter setup. This paper provides
useful guidelines to help the user make sensible choices.
\end{abstract}


\section{Introduction}\label{sec_intro}

Imaging surveys provide a general tool to access the average properties
of galaxy populations. A survey data set usually consists of an
arrangement of primary images in one or several filters. These data are
often accompanied by various supplementary data. Examples for such
surveys are COMBO-17 \citep{ref_combo}, DEEP1/DEEP2
\citep{ref_groth_strip}, GOODS \citep{ref_goods}, COSMOS
\citep{ref_cosmos} or the Hubble Ultra Deep Field \citep{ref_hudf}.

Common to all imaging surveys are the specific reduction methods
involved in the data analysis. After reducing the imaging data, which
normally consists of a mosaic of many potentially (partly) overlapping
tiles, scientific sources are detected and compiled in a source
catalogue. Depending on the scientific goals, more sophisticated
methods are then applied to analyse the morphology of the sources, \ie\
quantify the structure of their light-profiles. Finally, the resulting
additional structural parameters are added to the source catalogue.
Somewhere in this process the source catalogue might (optionally) get
cleaned from duplicate source entries or other artifacts.

For the main task, source detection and extraction, the code \sex\ by
\cite{ref_sex} has been widely-used in astronomy. Based on a
simple setup script \sex\ detects sources, estimates a background sky
level, measures primary shape information, like position, position
angle and axis ratio, and even performs aperture photometry. A key
feature is the ability to properly deblend close companion sources,
while at the same time avoid breaking single large sources up into
several pieces. Other features include a neural network to separate
stars and galaxies or the option to associate the detected objects with
a given list of input positions. \sex\ is designed with minimum user
interaction, support for large images and high execution speeds in
mind.

In order to analyse galaxy light profiles quantitatively, many codes
have been developed. The ones that are most widely used employ a
two-dimensional fitting method to model ellipsoidal radial profiles,
and include convolution with a point spread function (PSF).

One of these codes is \gimtwod, which was first employed as part of an
\iraf\ pipeline to analyse survey imaging data \citep{ref_gim2d}. Based
on a Metropolis algorithm to find the minimum in $\chi$-space,
\gimtwod\ mainly uses the \sersic\ profile \citep{ref_sersic}, which is
a general expression that includes both the de~Vaucouleurs and
exponential forms (see Sec.~\ref{sec_galfit} and eq.~\ref{eq_sersic}).
The minimisation method performs a global parameter space search. As a
result, \gimtwod\ is robust, however it requires large amounts of CPU
time compared to other codes \cite[\eg][]{ref_fitting}.

Another application for modelling light profiles is \budda\
\citep{ref_budda}. \budda\ was initially developed to perform
bulge/disc decomposition. However, it has recently been updated to
include also bar and central point source modelling. Moreover, it now
also features a double exponential profile for discs.

Finally, a rather versatile and effective method was presented by
\cite{ref_galfit, ref_galfit3}: \galfit. Like the aforementioned
programmes, it is a two-dimensional fitting code to extract structural
components from galaxy images. It is designed to model galaxies in as
flexible a manner as possible, by allowing the user to fit any number of
components and functional forms. \galfit\ therefore allows for the
possibility to not only fit simple situations, but also for fitting
more complicated setups including bulge, disk, bar, halo, etc.
This freedom has the major advantage that not only may the object of
prime interest be fitted, but so may the neighbouring sources -- at the
same time, as some situation may demand. Various light profile models
are built into the code, including the ``Nuker'' law \citep{ref_nuker},
the \sersic\ profile \citep{ref_sersic}, an exponential disc, Gaussian
or Moffat functions and even a pure PSF for modelling stars. \galfit\
convolves all model profiles, except for the PSF itself by the PSF to
simulate image smearing by Earth's atmosphere and telescope optics.

Although a scientist has a multitude of options to choose from for
fitting and detecting objects, analysing a complete survey to the end
of obtaining a source catalogue with galaxy parameters, requires many
intermediate steps. For example, duplicate sources from tile overlaps
have to be differentiated; the detection and fitting codes have to be
set up; a proper local background sky value has to be estimated;
resulting source parameters have to be compiled in a catalogue. As
these steps are fairly general we have built a code that simplifies all
these steps and largely automates the entire process. Our code, \gala,
performs all the required steps from a single setup and with minimal
manual interaction provides a fitting catalogue. It runs \sex\ to
detect sources and performs an automated \sersic\ fit using \galfit.
Amongst the various codes introduced above, we opted to use \galfit\
because it outperforms \gimtwod\ both in speed and reliability
\citep{ref_fitting} and allows a much wider range of light-profile
models than \budda. Upcoming versions will include additional features
like automated multi-component fitting. The code is available freely
for public download from our website at:
\url{http://astro-staff.uibk.ac.at/~m.barden/galapagos/}.

The layout of the paper is as follows. We start by giving an overview
of the structure of the code (Sec.~\ref{sec_structure}).
Then we elaborate on the methods involved in the individual components
(Sec.~\ref{sec_components}). Next, we present some fitting results
based on simulated data and provide details concerning the reliability
of the code (Sec.~\ref{sec_quality}). Subsequently, we give estimates
on the performance of \gala\ (Sec.~\ref{sec_performance}), followed by
a summary (Sec.~\ref{sec_summary}). Upon first reading this article we
suggest to skip Sec.~\ref{sec_components},
which address mainly the frequent \gala\ user. In the course of the
paper we assume a working knowledge of \sex\ and \galfit\ and refer the
reader to the publications by \cite{ref_sex} and \cite{ref_galfit}.

\begin{figure*}
\centering\includegraphics[width=12cm]{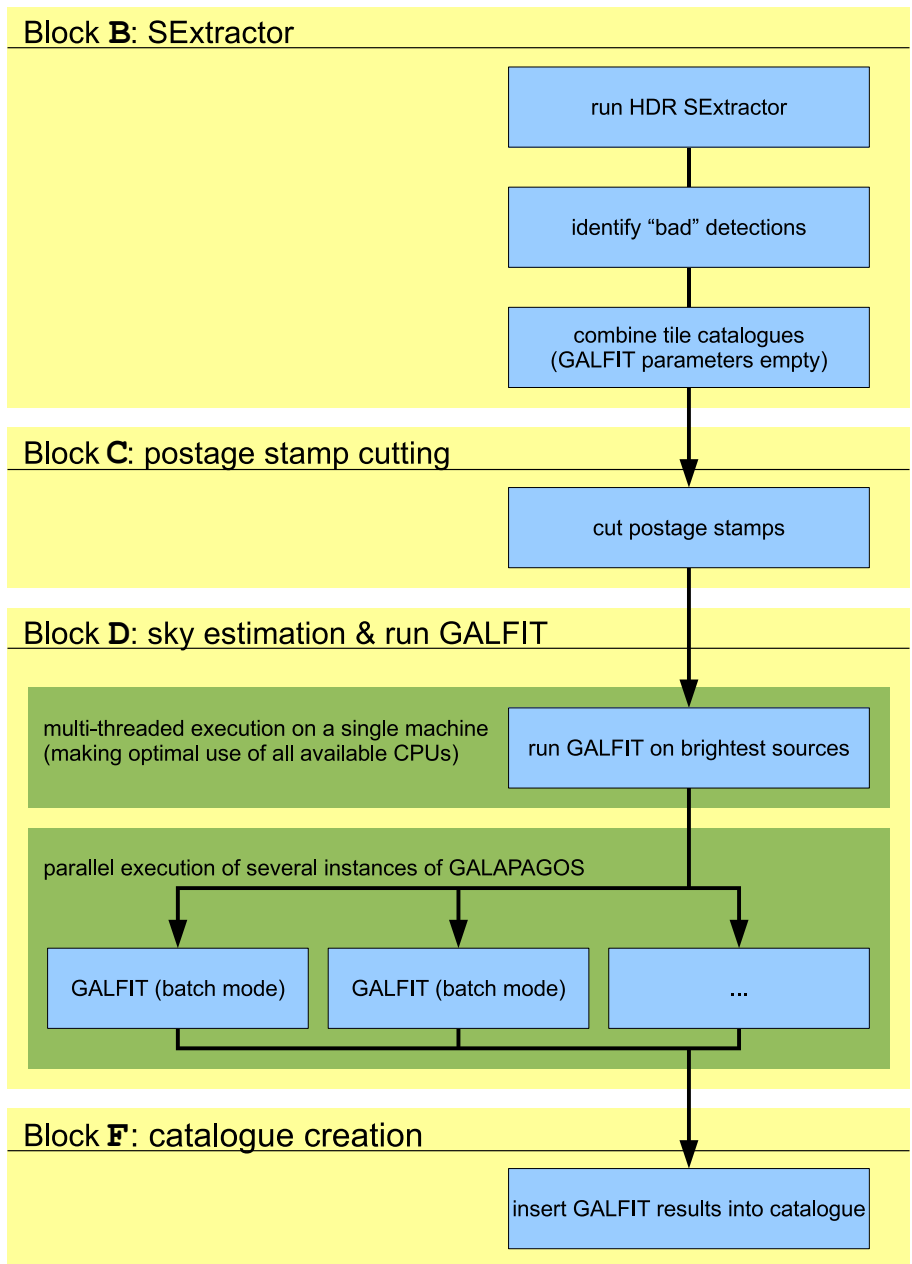}
\caption{Code structure. A yellow background indicates the four main
blocks, labelled according to the nomenclature of the \gala\ setup
file (see Fig.~\ref{fig_setupfile}). Fitting objects with \galfit\
(Block \texttt{(D)}) is a two-stage process (see
Sec.~\ref{sec_galfit}), which typically requires more than 90\%\ of
the total computation time (green background). We mark smaller tasks
by blue boxes. For further details see
Sec.~\ref{sec_structure}.}\label{fig_structure}
\end{figure*}

\section{Overview of Code Structure}\label{sec_structure}

\gala\ is divided into four main blocks, each of which is executable
independently from the others. This allows flexibility of repeating or
optimising certain segments of the analysis without re-running the
entire pipeline. These blocks are: \begin{enumerate}
 \item Detect sources by running \sex\ \verb|(B)|
 \item Cut out postage stamps for all detected objects \verb|(C)|
 \item Estimate sky background, prep. \& run \galfit\ \verb|(D)|
 \item Compile catalogue of all galaxies \verb|(F)| \end{enumerate}
Note that letters in brackets correspond to the respective sections in
the \gala\ setup file (see Sec.~\ref{sec_setup_main}). We visualise
this structure in Fig.~\ref{fig_structure}.

Also note that \gala\ does not create the PSF image, which is
required by \galfit\ in the fitting process. The user is responsible
for providing such an image. A proper PSF should have a sufficient
S/N, i.e. better than the brightest objects in the survey, in order not
to degrade the science data. Furthermore, it should contain all features
of the PSF down to the noise and it should not be truncated at the
edges. Also, it has to be background subtracted and normalised to a
total flux of 1.

\subsection{Source Detection}

In the first block \verb|(B)|, \sex\ is run to detect sources on the
individual survey images. Optionally, \gala\ features a high dynamic
range (HDR) mode for source extraction (Sec.~\ref{sec_sex}), which is
ideally suited for wide area and/or space-based, \eg\ \hst, data. After
a first pass, the user may refine this catalogue by identifying ``bad''
detections followed by re-running \sex. This may be required to fix
overly deblended sources or to remove spurious detections (see
Sec.~\ref{sec_bad_detections}). Once all tiles are analysed, \gala\
combines the individual output catalogues, rejecting duplicate sources
(see Sec.~\ref{sec_sex}) and optionally bad detections like cosmic rays
etc.~(see Sec.~\ref{sec_opt}).

\subsection{Postage Stamp Cutting}

To reduce the amount of time needed to ingest an image into \galfit\ it
is worthwhile to first extract each galaxy from the survey mosaic.
Therefore, in the second block \verb|(C)|, \gala\ estimates a size for
each object based on its \sex\ parameters. With this information it
computes the extent of a postage stamp. From the original survey
images, \gala\ then creates such a cutout for every object. It performs
the subsequent fitting with \galfit\ on these postage stamps (see
Sec.~\ref{sec_postages}). At this stage, \gala\ creates for every
survey image a ``sky-map'' containing information about the nature of
the pixel flux (either ``no flux'', ``sky'' or ``source''). It uses
this map later on to identify blank sky pixels (see
Sec.~\ref{sec_sky}).

\subsection{Sky Estimation and Fitting}\label{sec_structure_sky_fitting}

The third block \verb|(D)| performs the major fitting work. For every
object in the source catalogue it prepares and runs \galfit\ (see
Sec.~\ref{sec_galfit}). Accurate fitting analysis by \galfit\ requires
careful consideration, which includes identifying the proper sky
background, identifying neighbours and providing initial parameter
guesses to start the fitting.

\gala\ measures the sky using a flux growth curve including pixel
rejection based on the ``sky-map'', which was calculated in the
previous step (see Sec.~\ref{sec_sky}). It uses the full survey image
and not the small postage stamp to compute the sky. Note that even
though \galfit\ can fit the sky, \gala\ does not use this option to
avoid instances when neighbouring contamination makes accurate
determination infeasible, and to reduce the degree of freedom in the
fit. We provide further justification for and details on this approach
in Sec.~\ref{sec_postages} and Sec.~\ref{sec_sky}.

\subsection{Catalogue Creation}

In the last block \verb|(F)|, \gala\ reads the results of the fitting
from the headers of the \galfit\ output images and puts them into the
source catalogue (see Sec.~\ref{sec_cat}). Here, it removes a second
set of ``bad'' detections from the catalogue. Namely those that were
required in the fitting process to allow optimal results for
neighbouring objects. Usually, these are bright artefacts in close
proximity to relatively faint real sources (see
Sec.~\ref{sec_bad_detections}). Finally, \gala\ compiles the resulting
catalogue in a FITS-table.

\section{Components}\label{sec_components}

Subsequently, we describe in detail the methods involved in the
individual components of \gala. These include \sex\ and high dynamic
range (HDR) source extraction (Sec.~\ref{sec_sex}), compiling a
combined source catalogue (Sec.~\ref{sec_cat}), the cutting of postage
stamps (Sec.~\ref{sec_postages}), estimating a background sky level
robustly (Sec.~\ref{sec_sky}), and fitting with \galfit\
(Sec.~\ref{sec_galfit}). In the last part of this section we introduce
some technical mechanisms to optimise the code for robustness and speed
(Sec.~\ref{sec_opt}).

\begin{figure*}\centering\includegraphics[width=12cm]{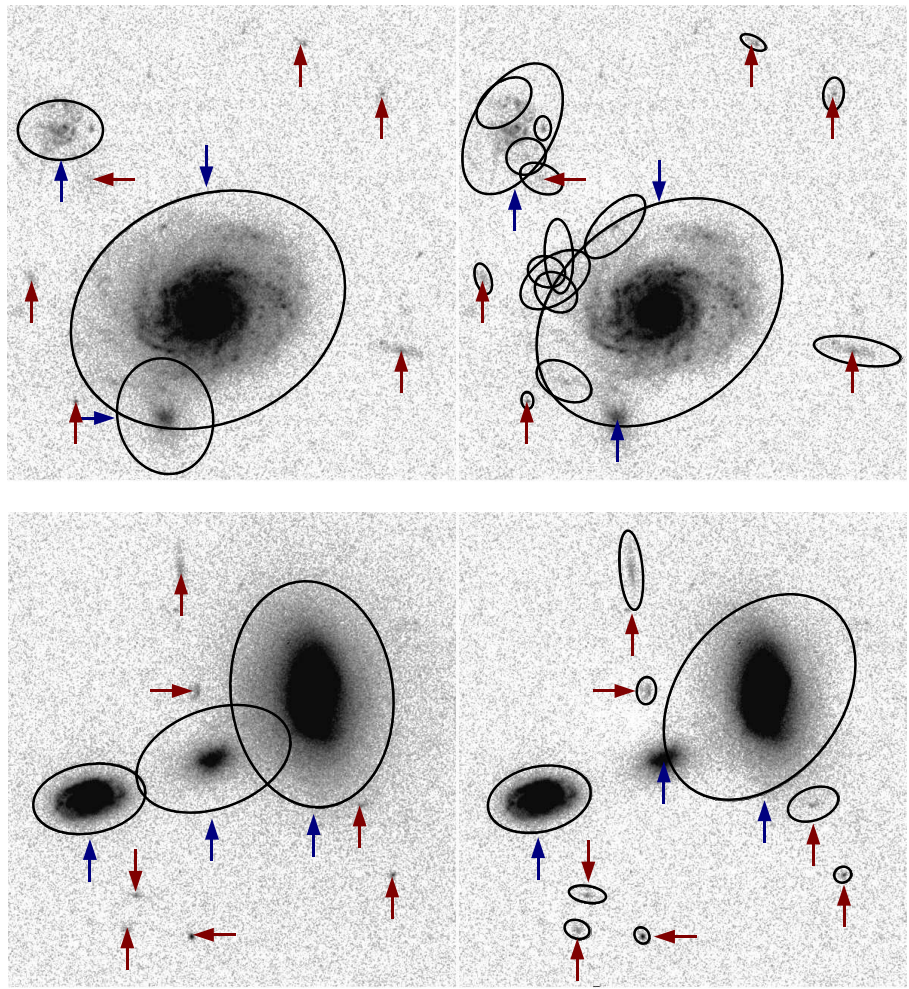}
\caption{Combining ``hot'' and ``cold'' \sex\ catalogues. The two
panels ({\it upper} and {\it lower}) show examples of a cold (left
side) and a hot (right side) SExtraction. Ellipses indicate the \sex\
Kron ellipses of the detected sources. Arrows mark objects from the hot
(red) and the cold (blue) catalogue that were incorporated into the
combined catalogue. Additional hot sources not marked with arrows (\eg\
in the upper right panel) were excluded from the combined catalogue as
described in detail in Sec.~\ref{sec_sex}.}\label{fig_hdr}
\end{figure*}

\subsection{SExtractor}\label{sec_sex}

\gala\ incorporates \sex\ to detect astronomical sources on individual
survey tiles. Details on how to operate \sex\ can be found in
\cite{ref_sex}. \sex\ uses the Kron radius to estimate the extent of a
galaxy \citep{ref_kron,ref_infante}. For both stars and galaxies, when
convolved with a Gaussian seeing, it encircles 90\% of their flux.
\gala\ applies the Kron radius \eg\ to estimate sizes of postage stamps
or to judge which pixels in an image are affected by light from
sources.

\sex\ has been used successfully with both ground- and space-based
data. Yet, recent large CCD arrays put the code to its limits due to
the wide range of object sizes and luminosities that are being observed
simultaneously. In classic pencil beam surveys, the objects of interest
are mostly faint and small. \sex\ is then fine-tuned to pick up such
sources properly at the cost of splitting up the occasional big bright
spiral galaxy into many pieces. On the other hand, wide area surveys
traditionally do not reach very deep. When fine-tuning \sex\ for these
surveys, emphasis is put on correctly deblending the larger and
brighter objects while losing some depth. In both applications one
reaches the dynamic range limit (in terms of object size and
brightness) and has to make a compromise of depth and proper
deblending.

\subsubsection*{HDR SExtraction}

Fortunately, there is a rather simple two-step approach using \sex\ to
overcome this problem. Firstly, one runs \sex\ in a so-called ``cold''
mode in which only the brightest sources are picked up and properly
deblended. As this will miss many faint sources, in a second setup
emphasis is put on depth. The second run we term ``hot'' mode.

Then one needs to combine the ``hot'' and the ``cold'' runs.
Firstly, all ``cold'' sources are imported into the output catalogue.
Then the Kron ellipses as provided by \sex\ of ``hot'' and ``cold''
sources are analysed. Every source position in the ``hot'' catalogue is
checked whether it falls inside a Kron ellipse of a ``cold'' source. If
it lies inside a Kron ellipse it is discarded and does not enter the
output catalogue; if its central position lies ``sufficiently'' outside
of all ``cold'' Kron ellipses it does enter the output catalogue.
``Sufficiently'' here refers to the possibility in \gala\ to
artificially enlarge the Kron ellipses slightly for this purpose:
parameter \texttt{B09} provides a scaling factor. Setting \texttt{B09}
to \eg\ 1.1 results in enlarging each Kron ellipse by 10\%.

In summary, it is important that the ``cold'' run properly
deblends all brighter objects, while the ``hot'' run is tuned to pick up
fainter sources. We term this mode ``High Dynamic Range (HDR)
SExtraction''.

To illustrate the process of including hot sources outside the Kron
radius of cold sources into a combined catalogue, see
Fig.\ref{fig_hdr}. In the upper left hand panel we show a ``cold'' run.
The big central spiral galaxy is deblended correctly with the fainter
galaxy below it. Also, the clumpy low surface brightness spiral in the
upper left corner is detected as a single source. All three sources are
taken over into the combined catalogue. Requiring a proper deblending
of the bright objects results in missing the faintest sources, though.
The ``hot'' run (upper right hand panel) picks those up. However, it
breaks the brighter galaxies up into many sources. In the
example, an
off-centre knot of the upper left galaxy was detected as a separate
object. Moreover, the outer regions of the central (and upper left)
galaxy are assigned separate source IDs. These ``spurious'' detections
change the effective size of the central galaxy (compare diameters of
the Kron ellipses in the left and right figures). Interestingly, the
relatively bright galaxy below the central object is not deblended
properly in the hot run. Furthermore, the size and position angle of
the upper left detection demonstrates the lower detection threshold of
the hot setup. In the hot setup a larger fraction of the low surface
brightness flux is included in the calculation of the position angle,
thus providing a much better estimate than the cold setup, which is
more heavily weighted towards the inner regions of the sources. Yet,
the values from the cold run enter the combined catalogue as deblending
is the more important source of error. Also, \galfit\ calculates
structural parameters like the position angle much more reliably. The
lower panels in Fig.~\ref{fig_hdr} show another example. Again, the
deblending in the hot run is bad, while in the cold run it is correct.
The faintest sources are only detected in the hot run. Bad deblending
in the hot run strongly affects the calculation of the position angle
of the brightest source, while in the cold run it is acceptable.

We developed and tested this method for the GEMS survey
(\citeauthor{ref_gems} \citeyear{ref_gems}; \citeauthor{ref_gems_cat}
\citeyear{ref_gems_cat}; for tests see \citeauthor{ref_fitting}
\citeyear{ref_fitting}). Subsequently, other major surveys have adopted
it as well, including COSMOS \citep{ref_cosmos_hst,ref_cosmos_lensing}
and STAGES \citep{ref_stages}. \gala\ provides the option for running
\sex\ in two-stage HDR or normal single-stage \sex\ configuration.

\begin{figure}\centering\resizebox{8cm}{!}{\includegraphics{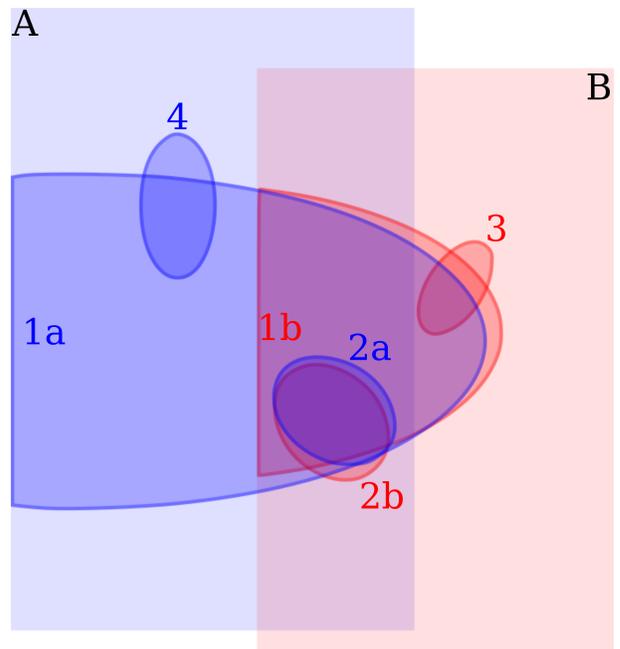}}
\caption{Combining \sex\ catalogues from neighbouring tiles. Tile A
contains sources 1a, 2a and 4, while objects 1b, 2b and 3 were detected
on tile B. Ellipses show the corresponding sizes from \sex. For a
description of what source ends up in the resulting table see
Sec.~\ref{sec_cat}.}\label{fig_combine_tiles} \end{figure}

\begin{figure}\centering\resizebox{8cm}{!}{\includegraphics{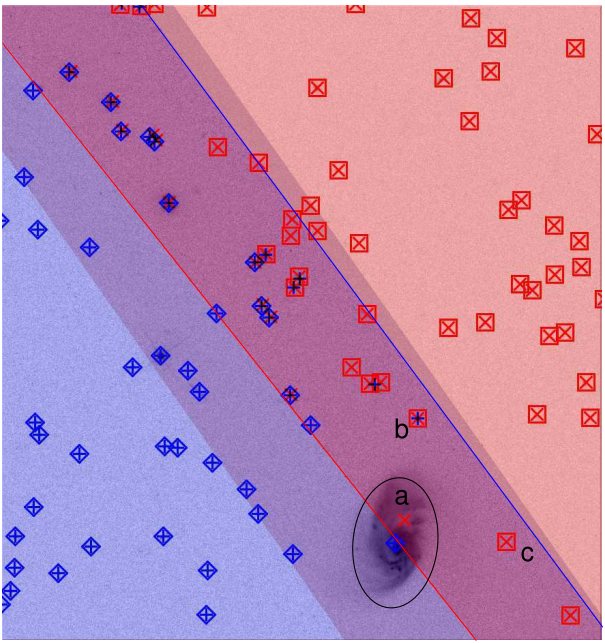}}
\caption{Combining \sex\ catalogues from neighbouring tiles. The left
image (blue area) extends out to the right (blue) diagonal line; the
right image (red area) extends out to the left (red) line. Shaded areas
outside of the lines corresponding to the respective image did not
receive sky flux. Pluses (blue) indicate source detections (hot and
cold already combined) from the left (blue) image; crosses (red) mark
detections from the right (red) image. Diamonds (blue) highlight
objects that are contained in the combined catalogue if they originated
from the left (blue) image; boxes (red) highlight those that were taken
from the right (red) image. Catalogue combination is based on the \sex\
ellipses (see Sec.~\ref{sec_cat}). Such an ellipse is shown in case a).
The source from the right image (red) is rejected as it lies inside an
ellipse of a detection in the left image (blue), which is further away
from the respective image boundary (blue and red lines). For the same
reason in case b) the red source is kept. The detection in case c) does
not even have a counterpart in the other catalogue.}\label{fig_comb1}
\end{figure}

\subsection{Catalogue Compilation}\label{sec_cat}

Compiling the output source catalogue is a two-stage process. \gala\
creates a first combined catalogue from the \sex\ output tables. In the
subsequent model fitting process, \gala\ fills this catalogue with the
\galfit\ output parameters.

When putting together catalogues from potentially overlapping images,
\gala\ has to take care of removing detections of the same source on
multiple images. To this end, it uses the world coordinate system of
the images to translate pixel coordinates from one to another image.
Next, \gala\ calculates the distance to the image border for each
source (not only those in the overlap area) in the corresponding image
catalogues. The area containing flux (pixels with non-zero values)
defines the image border. This is crucial in particular for
non-rectangular images (\eg\ from \hst). Now, \gala\ sorts the two
catalogues by border-distance. It starts with the source farthest
from the edge, which we assume to be on image {\it A} (source 1a;
see Fig.~\ref{fig_combine_tiles}). Then it checks whether there are
sources inside the Kron ellipse of the current object in the
neighbouring image {\it B} (sources 1b, 2b and 3). If it finds any such
targets, \gala\ removes them from the list. Note that \gala\ does not
remove objects overlapping with the source from image {\it A} from the
list (sources 2a and 4). Following this scheme it works through the
complete list, from the farthest to the closest objects to the
boundary, and constantly updates the list in the process.

A problem
arises for sources, say in image {\it A} (source 1a), extending over a
radius larger than the size of the overlap area and having overlapping
detections on image {\it B}, which are not covered by image {\it A}
(source 3 in Fig.~\ref{fig_combine_tiles}). Or put differently,
if source 1 is \eg\ deblended differently in image {\it A} than in image
{\it B}, sources might get lost in the combination process. In such a
case, \gala\ includes the main source from image {\it A} (source 1a) in
the catalogue and all overlapping sources from image {\it A} (sources 2a
and 4). Overlapping sources from image {\it B} it removes, though
(sources 2b and 3). However, in cases where source 1 was
over-deblended in image {\it B}, but not in image {\it A}, this would
result in a welcome clean-up of the catalogue by removing the spurious
source 3. Although this problem cannot be unambiguously solved, in
practice it rarely occurs. It can be avoided completely if the largest
source in the survey is smaller than the overlap between survey images.

Fig.~\ref{fig_comb1} shows an example for this procedure to remove
duplicate detections. The bright galaxy a) is just on the edge of the
red image. As only half its flux is visible on that image, the
calculated centre is far off from the real position. The blue image
fully contains this galaxy. The two central positions (in red and blue)
being so different, a normal nearest neighbour matching algorithm with
a maximum matching radius would not have been able to identify the two
detections as the same source. In the proposed scheme, though, the red
detection is not put into the combined catalogue, for being inside the
Kron ellipse of a source that is further from the image edge in the
blue image. Similarly, the red source b) is further from the image edge
than the blue source and thus, we reject the blue object. Objects
without counterpart in the other image, as in c), we do keep in the
combined catalogue.

As \gala\ performs duplicate removal before running \galfit, it does
not fit sources twice. Note that for fitting sources at the edge of an
image, \gala\ takes objects on neighbouring survey images into account
as well (see Sec.~\ref{sec_opt}).

\subsection{Postage Stamps}\label{sec_postages}

To optimise galaxy fitting with \galfit, \gala\ cuts the science images
into smaller sections centred on individual sources. The advantage of
using such postage stamps is that the total fitting time and the
demand on main memory can be reduced. Even rather deep optical surveys
contain large fractions of empty sky, which can mostly be excluded from
the fit once the information from the sky pixels is effectively used to
estimate the background (see Sec~\ref{sec_sky}; even masking cannot
totally diminish this advantage). In a typical one-orbit \hst\ survey
around a factor of 2 in the total number of pixels can be saved.
Moreover, although \galfit\ allows simultaneous fitting of multiple
sources, modelling more than a handful of objects at the same time
quickly becomes rather impractical. Thus, to optimise automated fitting
of large numbers of sources, \gala\ incorporates a postage stamp
cutting facility.

To determine the size of the postage stamps, \gala\ uses the Kron
radius. The user specifies a scale factor \verb|C03| by which the Kron
radius is enlarged. The decision for this scaling should be guided by
trying to find a compromise between maximal area, to include as much
flux of the central source (and maybe the closest neighbours) as
possible, and minimal area, to speed up computation time of \galfit.
Finding a good compromise is important as elliptical galaxies require a
larger area than spiral galaxies, owing to their extended and slowly
dimming, low surface brightness wings. For the one-orbit \hst\ surveys
GEMS and STAGES, we found a factor of 2.5 to work well. \gala\ does not
enlarge the size of the postage stamps in the presence of close
neighbouring galaxies. However, they are properly taken into account in
the fitting process (see Sec.~\ref{sec_galfit}).

We note that a disadvantage of using postage stamps for fitting {\it
with the background sky as a free fit parameter} (which we discourage
the user from in the context of \gala) is that the fit results will be
biased if the postage stamp does not contain enough empty sky pixels. In
such a case the $\chi^2$ of the fit might indicate a good fit, yet the
result would be flawed by attributing too much or too little flux to the
object. This could potentially also have a strong impact on other
structural parameters. Therefore, \gala\ does {\it not} allow a free fit
of the sky background within \galfit, but estimates a value before the
fitting. We give details on the background estimation in the following
section.

\begin{figure*}\centering\includegraphics[width=17.5cm]{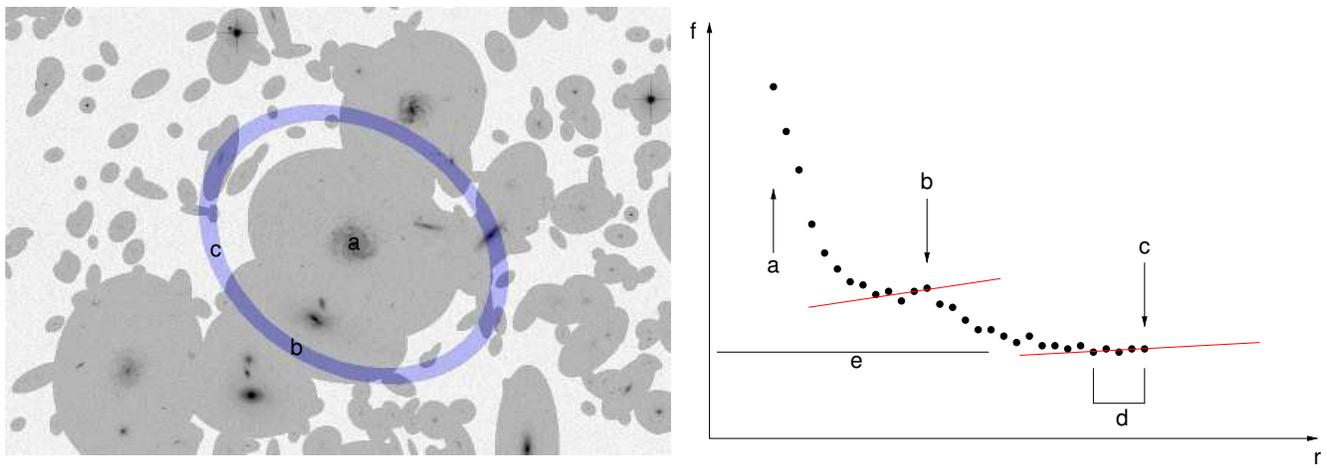}
\caption{Sky estimation. {\it Left:} The average flux $f$ measured in
elliptical annuli (blue) centred on an object (here {\it a}) determines
the background level. In each annulus, we exclude regions surrounding
other sources from the calculation (shaded area). For the indicated
annulus, we exclude dark blue shaded regions {\it b} -- only light blue
regions {\it c} define the average background flux. {\it Right:} Flux
$f$ measured in an elliptical annulus as a function of radius $r$. {\it
a:} Starting radius. {\it b:} Slope (indicated by the diagonal lines)
turns positive for the first time, \eg\ due to galactic structure at
large radii. {\it c:} Slope turns positive for the second time. Here we
stop the iteration. {\it d:} We compute slope measurements from the
last $n$ sky estimates (here: n=5; $n$ is a user parameter). {\it e:}
The adopted background sky level. See Sec.~\ref{sec_sky} for
details.}\label{fig_sky} \end{figure*}

\begin{figure*}\centering\includegraphics[width=17.5cm]{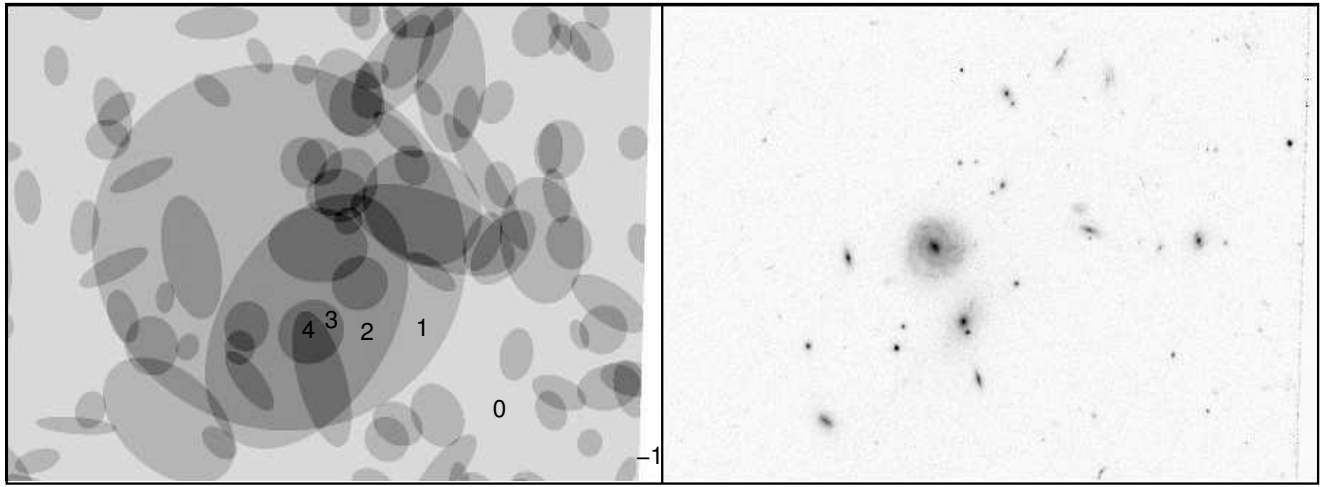}
\caption{The ``skymap'' ({\it left}): for each object that was detected
in the image ({\it right}) we calculate the Kron ellipse and scale it
up. Pixels inside a Kron ellipse get the value one. Pixel values stack,
\eg\ where two Kron ellipses overlap, the pixel value is two. Blank sky
has a value of zero; pixels without astronomical flux, as occurs after
removing image distortions \eg\ in \hst\ images, have a value of -1.
Some pixel values are indicated.}\label{fig_skymap} \end{figure*}

\subsection{Sky Estimation}\label{sec_sky}

Obtaining a precise sky level is the most critical systematic in galaxy
surface brightness profile fitting \citep[see
\eg][]{ref_dejong,ref_fitting}. To obtain a precise background
measurement \galfit\ is capable of including the sky as a free
parameter when fitting a celestial source. However, using the sky as a
free parameter requires an appropriate size of the input image, \ie\ it
has to contain {\it all} the flux of the primary source and most of the
flux of neighbouring sources that are to be fitted simultaneously and
ample sky. For estimating a proper sky background, the image should be
as large as possible. However, as detailed above, large postage stamps
become impractical once too many neighbouring sources are included.
Only a manual setup may allow using the sky as a free model parameter.
To enable automated processing of large numbers of objects, \gala\
incorporates its own subroutine to obtain an optimal sky measurement
before running \galfit\ and hence uses a fixed value during fitting.
With the proper setup, the resulting \gala\ estimate improves
significantly over values obtained from \sex.

We use a flux growth method to estimate the local sky around an object.
Calculating the average flux in elliptical annuli centred on the object
of interest while excluding other sources or image defects, we obtain
the background flux as a function of radius. Once the slope over the
last few measurements levels off, \gala\ stops and determines the sky
from those last few annuli (see Fig.~\ref{fig_sky}).

\begin{figure*}\centering\includegraphics[width=17.5cm]{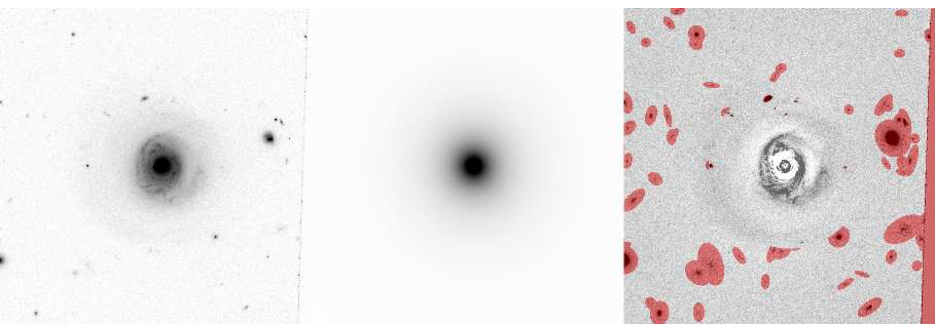}
\caption{Fitting \sersic\ profiles with \galfit. From left to right,
the panels show the original galaxy image, the \sersic\ model and the
residual of image and model, respectively. \gala\ excludes (masks)
areas shaded in red from the fit. In this example no bright secondary
sources were detected. The next brightest object after the primary
source is too far away to become a secondary (for details on the
definition of primary and secondary sources see Sec.~\ref{sec_galfit}).
Note that the masked region at the right edge of the image results from
the irregular shape of the \hst\ images. This area has not received any
flux and thus \gala\ masks it as well.}\label{fig_galfit} \end{figure*}

For this procedure to work, we create a ``sky-map'', \ie\ a copy of the
input images where the pixel values indicate the nature of the
contained flux. In the sky-map a pixel value of 0 stands for blank
background sky, while positive numbers indicate the presence of a
source. A value of -1 indicates no flux at all, as happens with \hst\
images that are geometrically distorted (see Fig.~\ref{fig_skymap}).
One might think that to make the decision between source or sky the
\sex\ segmentation map \citep[for a definition see][]{ref_sex} might
suffice. Unfortunately, the level out to which \sex\ detects objects is
rather limited. In particular with elliptical galaxies \sex\
underestimates the flux belonging to the object significantly. Changing
the \sex\ setup parameters cannot totally remedy this. Therefore, a
significant number of pixels still containing some source flux would be
assigned as ``sky''. To circumvent this problem, we instead use Kron
ellipses to determine the extent of an object. \gala\ regards any pixel
inside \verb|D03|$\times R_\mathrm{Kron} + $\verb|D05| as containing
source flux (for STAGES: \verb|D03| $=3$; \verb|D05| $=20$~pix). Note
that this scaling factor \verb|D03| does not have to coincide with the
scale for the size of the postage stamps \verb|C03|. Note also that
\gala\ records the total number of objects that might contribute to a
certain pixel, \ie\ when \eg\ two sources overlap, the value in the
intersection of the two Kron ellipses is also two. A weight map
(exposure time map) specified by the user defines the off-chip pixels,
which are given a value of -1. \gala\ identifies these as pixels with
zero exposure time.

\gala\ takes special care to minimise the impact of large nearby
sources on the background estimation process for the current object. To
that end, \gala\ relies on the \sex\ output catalogue to provide shape
information. Under the assumption that all sources have a \sersic\
index $n=4$ and a half-light radius
$r_e=\left(\tt{flux\_radius}\right)^{\alpha}$, with the \sex\ catalogue
parameter {\tt flux\_radius} and a user specified power $\alpha$ (we
chose $\alpha = 1.4$; \verb|D11|) to convert the \sex\ {\tt
flux\_radius} to a ``true'' half-light radius, \gala\ calculates the
flux of all catalogue objects at the position of the current source.
Any source exceeding a user specified limit \verb|D09|, \gala\ regards
as an important flux contributor for the current object. Subsequently,
we will term the sources that are selected that way ``contributors''.
Note that \verb|D09| has the units of a magnitude, \ie\ ``exceeding''
the given limit implies a number smaller than this value. As the \sex\
{\tt flux\_radius} is a rather poor proxy for the true half-light
radius and without proper estimate for the \sersic\ index, we opt for a
rather conservative limit of this flux cut.

If a proper \galfit\ fit exists for the contributors, \gala\ subtracts
their model profile from the input image temporarily, \ie\ for the time
of the current background estimation. Note that removal of a model
profile includes convolution with the telescope PSF before subtraction.
In order to optimise the profile subtraction, \gala\ processes the
\sex\ source catalogue in order of increasing magnitude. As the very
few brightest sources have a significant impact on both the sky
estimation and fitting of a large number of fainter sources, starting
the fitting process with the brightest galaxies is essential. We give
further details about the sorting process in Sec.~\ref{sec_opt}.

Normally, the Kron ellipse of the current object defines the starting
radius for the iterative measurement of the sky background in
increasing annuli. In case of the presence of potentially dominant flux
contributors, for which no \galfit\ model exists yet, and hence were
not subtracted from the input image, \gala\ increases the starting
radius to the maximal distance of all such sources from the current,
as they might potentially influence the fitting.
For each sky annulus, it estimates an average flux value excluding any
pixels that were flagged as containing an object (or that were flagged
as having a defect or no flux) in the skymap. Firstly, of the
distribution of the remaining pixels, \gala\ symmetrically clips all
$3\sigma$ outliers. Then it fits a Gaussian function to the leftover
distribution, producing a mean value for the current annulus. After
each new sky annulus measurement, \gala\ calculates a robust linear fit
to the last few estimates (\verb|D13|; in the case of STAGES 15
measurements). As long as source flux is still measurable, this slope
is negative. Once this process reaches the true background, the
estimated slope should start to randomly change its sign. When this
happens for the second time, \gala\ stops the loop and obtains the
final background value from the last \verb|D13| measurements. Stopping
the process at the first positive slope measurement often results in
suboptimal estimates as galactic inhomogeneities (like spiral arms)
might produce dips sufficient to produce a slope sign change. However,
using a much later slope change (than two) in practice is not necessary.
Note that neighbouring sources are not a problem for the termination of
this iteration as the method takes special care to take their influence
into account (as shown above). This whole process is fully
user-configurable, including options for the width of the sky annuli
\verb|D07|, their spacing \verb|D06|, the initial starting radius
\verb|D08| and the magnitude cut \verb|D09|.

\subsection{GALFIT}\label{sec_galfit}

Of the various light profiles built into \galfit, the most general one
for galaxy fitting is the \sersic\ model. It is also used by \gala:
\begin{equation}\label{eq_sersic}
\Sigma\left(R\right)=\Sigma_e\cdot\exp\left(-\kappa\left[
\left(R/R_e\right)^{1/n}-1\right]\right),
\end{equation}
where $R_e$ is the effective or half-light radius, $\Sigma_e$ is the
effective surface brightness, $\Sigma\left(R\right)$ is the surface
brightness as a function of radius $R$, $n$ is the \sersic\ index and
$\kappa=\kappa\left(n\right)$ is a normalisation constant. The \sersic\
profile is a generalisation of a de~Vaucouleurs profile with variable
\sersic\ index $n$. An exponential profile has $n=1$ while a
de~Vaucouleurs profile has $n=4$.

A simple setup script controls profile modelling with \galfit. It
contains information about input and output file locations, PSF image,
bad pixel mask, etc. A list of starting guesses defines what
light-profiles are to be fitted. Although the downhill gradient method
incorporated in \galfit\ is often speculated to be prone to converging
to a local instead of the global minimum, in practice we find it to be
extremely robust, even in comparison to global parameter space search
algorithms \citep{ref_fitting}. In application to high redshift survey
data, the other two noteworthy features are the included bad pixel mask
(\ie\ pixels that are excluded from the fitting) and the number of
simultaneously fitted objects.

We show an example of \galfit\ output in Fig.~\ref{fig_galfit}. The
left image presents the input postage stamp. In this case a single
component (one object) was fitted. We show the resulting \sersic\ model
in the middle. Note that the brightness cuts and scaling in both left
and middle panels are the same. The right panel displays the difference
image of input minus model. Bright spiral features and dark dust lanes
that strongly deviate from the smooth \sersic\ profile are very
prominent in this image. In order not to bias the fit by neighbouring
sources and the image boundaries, \gala\ excludes the shaded region
from the fit by applying a bad pixel mask (see below).

In order to define which objects do not have a high importance for the
current fit and hence may be masked instead of being fitted
simultaneously, we define the following terminology: the target for the
current fit is the {\it primary source}; any object whose expanded Kron
ellipse overlaps with that of the primary are {\it secondary sources};
objects without any overlap with the primary we term {\it tertiary
sources}. We consider tertiary sources not to be important for the
quality of the fit. As a result, we mask and exclude them from the
modelling (each pixel in a mask image is ignored during the fit by
\galfit). Secondary sources might have an impact on the outcome of the
parameters of the primary. Therefore, we fit them simultaneously
with the primary. We treat contributors (for a definition see
Sec.~\ref{sec_sky}) as secondaries. The difference between
contributors and secondaries is, that we do not require an overlap of
the Kron ellipses for contributors.

\begin{figure}\centering
\resizebox{8cm}{!}{\includegraphics{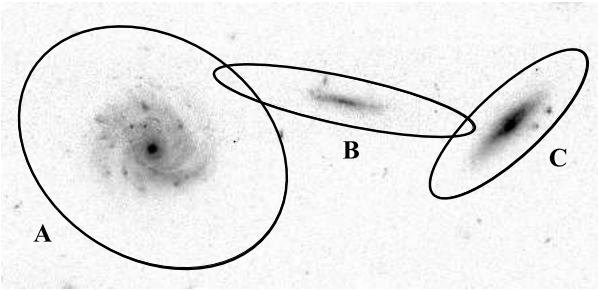}}
\caption{Optimisation of the \galfit\ setup. Circles indicate the Kron
ellipses used for classifying the detected objects (as secondaries or
tertiaries). This example was taken from real data. However, for
clarity we do not mark the faintest detections in this image. For
details see Sec.~\ref{sec_galfit}.} \label{fig_fix} \end{figure}

\begin{figure*}\centering\includegraphics[width=17.5cm]{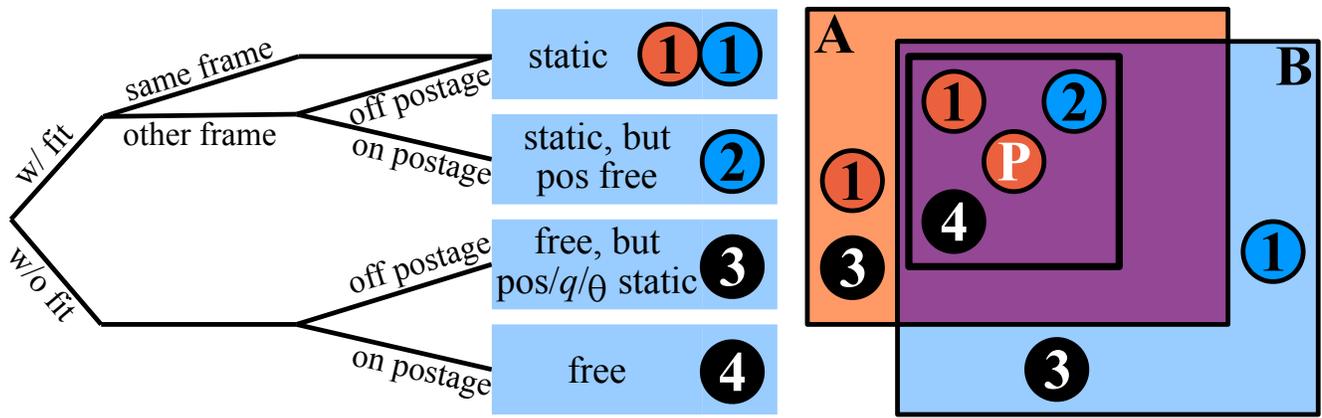}
\caption{\galfit\ parameter setup scheme for secondaries and
contributors. Depending on the relative position to the primary target,
\ie\ on the same postage stamp or not, with a pre-existing fit from the
same or another survey image or without a fit, we show the setup for
the \galfit\ parameters ({\it left panel}): static implies all
parameters are fixed to their initial guess (\ie\ the \sex\ estimates),
while free means that they are variable throughout the fit. In some
cases the position pos, the axis ratio $q$ or the position angle
$\theta$ take a different state than the remaining fit parameters. We
visualise the situation in the {\it right panel}: The current primary
{\it P} is located on the red survey image {\it A}, which has some
overlap (purple) with the blue image {\it B}. The solid black outline
indicates the postage stamp corresponding to {\it P}. Potential
secondaries or contributors are numbered. For sources with a black
background ({\it 3} \& {\it 4}) no prior fit exists (\sex\ values are
used as static/fixed profile parameters), while for targets shown
either in red ({\it 1}) or blue colour ({\it 1} \& {\it 2}) a fit from
the respective survey image is available. For further details see
Sec.~\ref{sec_galfit}.}\label{fig_simultaneous} \end{figure*}

\subsubsection*{Simultaneous Fits}

Often, sources are so close to each other, that they are best fitted
simultaneously. One might argue that after a simultaneous fit of two
sources, {\it A} and {\it B}, the best parameters are known for both
objects. However, in general this is not true. For example, we construct
a situation with three sources {\it A}, {\it B} and {\it C}, where {\it
C} is on the opposite side of {\it A} with {\it B} being in the middle
(see Fig.~\ref{fig_fix}). Let {\it A} be the brightest source of the
three, \ie\ {\it A} is fitted firstly (see also Sec.~\ref{sec_opt}).
The Kron ellipses of {\it A} and {\it B} and those of {\it B} and {\it
C} overlap, while the Kron ellipses of {\it A} and {\it C} do not.
Fitting of {\it A} implies fitting of {\it B} simultaneously: {\it B}
is a secondary to the primary {\it A}. {\it C} is not connected to {\it
A} and therefore following our prescription we mask it. As a tertiary
we exclude it from the fit. The resulting fit for {\it B} thus is not
optimal, as it neglects the presence of {\it C}, which is important for
fitting {\it B}, but not for fitting {\it A}. To obtain the optimal fit
for {\it B}, we have to fit {\it B} as a primary. In this case {\it A}
and {\it C} are secondaries as their Kron ellipses both overlap with
{\it B}, and are fitted simultaneously. To speed up the fitting of {\it
B}, we can now insert the known parameters for object {\it A}, thus
effectively removing one component from the fit. This example
highlights the importance of fitting all objects once as primaries,
while secondaries may be made static if a fit already exists.

Normally secondary sources are fitted simultaneously with the current
primary object (see above). Using a pre-existing fit (as in the
example) as static parameters for a secondary source, thus, is an
exception to this rule. A further complication is that the existing fit
for the secondary may have been obtained from a different survey image
as the current primary. In that case, the central position of the
secondary has to be converted via the world coordinate system
information from the original pixel coordinates to the current system
of the primary. Therefore, to allow optimal centring after such a
conversion, \gala\ fixes all parameters for the secondary, but its
central coordinates. If in the previous example, when fitting object
{\it B} with a pre-existing fit of source {\it A}, the fit for {\it A}
was performed on a different survey image than {\it B}, then the pixel
centre of {\it A} would not be static. However, a free pixel centre is
only required if the centre of the secondary {\it A} is also inside the
postage stamp of the current primary {\it B}. If the centre is off the
postage stamp, sub-pixel accuracy is not required any more for an
optimal fit, and all components of {\it A} are made static. We
visualise this situation in Fig.~\ref{fig_simultaneous} (case {\it 1}
and {\it 2}).

Furthermore, if no fit exists for a secondary, a free fit for that
source is not always the best solution. In the case that the centre of
the secondary is not on the postage stamp, a free fit results in too
many degrees of freedom. In \gala\ we opt to then fix the position,
axis ratio and position angle to the values provided by \sex\ (while
leaving the \sersic\ index $n$ and the half-light radius $R_\mathrm{e}$
as free parameters; see Fig.~\ref{fig_simultaneous} case {\it 3} and
{\it 4}). This is justified because,
on one hand, more than half the flux of the secondary cannot be seen by
\galfit, thus making it increasingly difficult to come up with precise
estimates for these parameters. On the other hand, the values given by
\sex\ usually have high enough accuracy not to bias the fit of the
primary significantly.

In addition to the ``normal'' sources (secondaries and tertiaries) in
the immediate surroundings of the current object, \gala\ has to take
bright and large contributors as defined in Sec.~\ref{sec_sky} into
account as well, although these sources may be off the current survey
image. It treats them as secondaries without the requirement of their
Kron ellipse to overlap with the Kron ellipse of the primary. In terms
of the parameter setup, \gala\ handles them exactly like other
secondaries.

\begin{figure*}\centering\includegraphics[width=12cm]{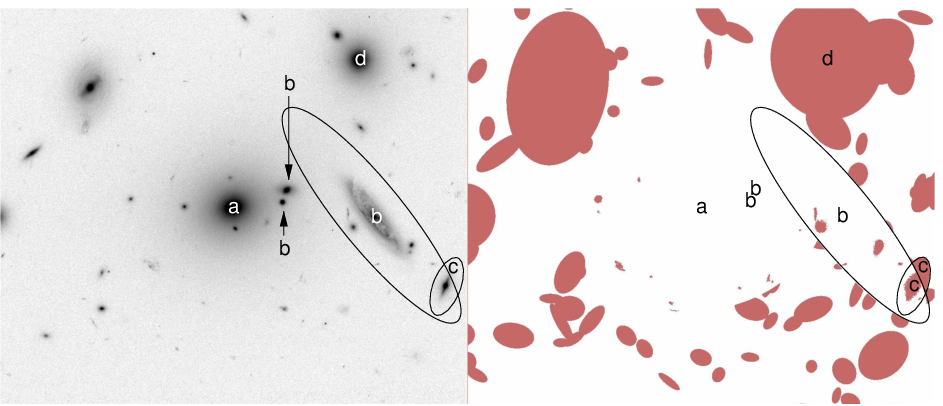}
\includegraphics[width=12cm]{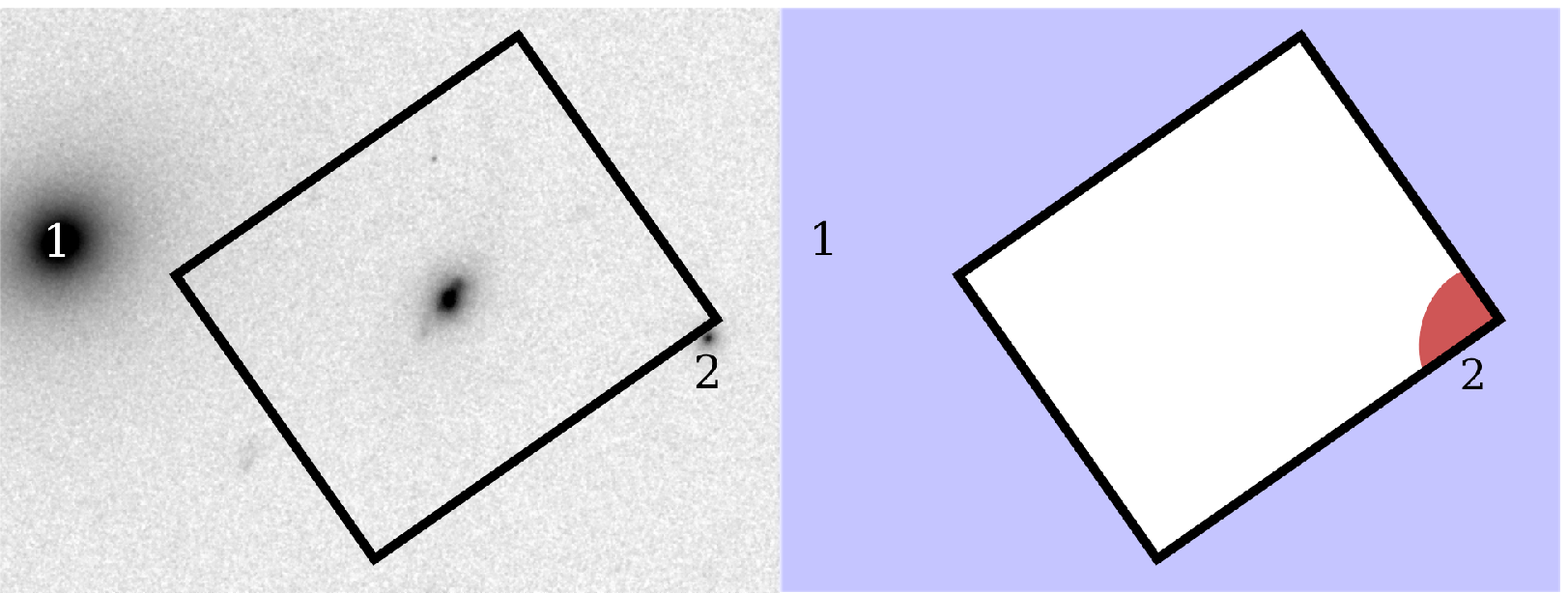} \caption{Mask creation. The
left panels show galaxy images; the right panels the corresponding bad
pixel masks. In the bad pixel masks white and red represent good and
bad pixels, respectively. {\it Upper panels:} a) and b) indicate
primary and secondary sources, respectively. c) and d) mark examples of
tertiary sources: c) is only partly masked, as it overlaps with a
secondary source; d) is masked completely for not having any overlap
with the primary or a secondary. {\it Lower panels:} The plotted area
shows a postage stamp (indicated by the solid rectangles) and some of
its surroundings. Note that the postage stamp is tilted (representation
in world and not pixel coordinates) and that the blue area is actually
not part of the postage stamp. 1) is a source that might potentially
contribute to the fit of the primary, due to its brightness, and is
included as a secondary source (with parameters fixed from a previous
fit) although its centre is off the current postage stamp. 2) is a
tertiary source without overlap with the primary and being too faint to
contribute significantly, and is therefore masked completely (red
pixels inside the postage stamp).}\label{fig_mask} \end{figure*}

\subsubsection*{Bad Pixel Masks}

\galfit\ supports so-called bad pixel masks \citep[see][]{ref_galfit}
to exclude image regions from fitting and thus speed up the fitting
process. As tertiary sources may overlap with secondaries, we take the
following approach to define the area to be masked. In general, \gala\
masks the full Kron ellipse, enlarged by a user-specified factor
\verb|D04| (which may have a different value as the one used for
computing the skymap \verb|D03|) and an additional offset \verb|D05|,
for the fitting. If the Kron ellipse of the tertiary overlaps with the
Kron ellipse of the secondary, \gala\ {\it includes} the intersection
in the fit. However, as the included area might contain significant
flux (maybe even the nucleus) of the tertiary, it {\it excludes} any
pixel marked in the \sex\ segmentation map as belonging to the
tertiary. Thus, the resulting shape of the mask may look complicated,
yet this procedure ensures having the fit of the secondary targets only
mildly affected. The primary source should not be significantly
affected at all.

To speed up the fitting process by reducing the number of simultaneous
fits, \gala\ masks secondary objects based on a magnitude criterion
\verb|D16| (for extended and \verb|D17| for point sources) in
comparison to the primary source. In the case that they are too faint
compared to the primary, \gala\ ``downgrades'' them to tertiary status
and treats them as such, \ie\ it masks their Kron ellipses completely,
but for parts which overlap with other secondaries or the primary that
are not covered by the \sex\ segmentation map.

\gala\ also masks pixels that have a value of zero in the weight map,
\ie\ an exposure time of zero. Obeying these rules results in masks as
shown in Fig.~\ref{fig_mask}.

\subsubsection*{Parameter Constraints}

\galfit\ not only applies a bad pixel mask, but also allows
fit parameters to be constrained in various ways. Examples
are keeping a parameter within an acceptable absolute range
(\eg~\sersic\ indices should satisfy $0.5<n<8$) or a relative range
depending on the given input values. Parameters might even be
constrained with respect to each other or other components. For more
details see the GALFIT homepage
\url{http://users.obs.carnegiescience.edu/peng/work/} \url{galfit/galfit.html}.

With respect to \sersic\ fitting in \gala\ providing a suitable range for
the \sersic\ index $n$ and the half-light radius $R_\mathrm{e}$ has a
stabilising effect on the procedure. To this end in \gala\ a limit
on the relative difference between \galfit\ and \sex\ magnitude is
imposed as well. \gala\ incorporates global constraints on the \sersic\
index $n$ ($0.2<n<8.0$), the half-light radius $R_\mathrm{e}$
($0.3<R_\mathrm{e}<$ \verb|E11|) and the fit magnitude $m$
(\verb|E12| $<m_\mathrm{GALFIT}-m_\mathrm{SExtractor}<$ \verb|E13|).

\subsection{Computational Optimisation}\label{sec_opt}

In the following section we will describe additional characteristics of \gala\
that increase the efficiency and robustness of the code.

\subsubsection*{Sorting and Parallel Computation}

After running \sex\ and cutting postage stamps, \galfit\ fits the
individual catalogue sources. Because efficient removal of brighter
sources is needed for accurate estimation of the sky background, an
ordered processing is required. This is extremely inefficient in terms
of total CPU time, we have developed methods to speed up this sequential
process. In the next paragraphs we will describe the mechanisms that are
incorporated into \gala\ to switch from sequential to parallel
processing and to increase the overall efficiency and robustness of the
code.

To optimise the execution time, \gala\ performs fitting in a
rank-ordered sequence starting with the brightest source in the survey
and progressing to the fainter ones. The advantage of this procedure is
twofold:\\
$\bullet$ Faint neighbours of bright sources do not have to
be included in a simultaneous fit (as a second component), as they do
not influence the resulting fit parameters of the bright object
significantly. The magnitude difference between faint and bright
neighbours is a free user parameter (\verb|D16|, \verb|D17| if the
primary is a galaxy or a star, respectively).\\
$\bullet$ When a faint
source has a brighter neighbour, which has to be included in the fit as
well, parameters for that object will already exist from a previous
fit. Hence the variables for that component can safely be held fixed to
the best values. This reduces the total number of degrees of freedom
and increases the computation speed for a large number of sources
tremendously. Another reason in favour of
sorting objects by magnitude is that the efficiency with which bright
contributors are included in the current fit is greatly enhanced.

The weakness of the sequential approach is that it voids the speed
benefits of parallel processing. To alleviate this problem, we devise
two methods:\\ a) Consecutive sources in a rank-ordered list are
usually sufficiently far apart to not affect one another (the average
object size is much smaller than their typical distance). Therefore,
\gala\ starts the next object in the sequence as a new process on
another CPU (core), given that its distance from other sources in the
queue is large enough (\verb|D20|). The extent of the brightest object
in the survey determines this distance and it should exceed the limit
out to which this object might have an influence on the fitting of
neighbours. If the next source in the queue is too close to objects
currently being executed, the code waits for these objects to finish.

b) Generally it is possible to parallelise the analysis by running the
code on one survey image only, at a time, by encapsulating the sky
fitting and \galfit\ processing. This will then enable the user to run
several instances of the code simultaneously on $n$ different
computers, thus reducing total computation time by a factor of $n$.
This is realised in \gala\ by specifying which tiles are to be
processed in a so-called ``batch file'' (\verb|E01|). The problem with
this approach is that sources may extend from one survey image onto the
next. Therefore, one might run into the situation where tile {\it A} is
fitted before tile {\it B}, with the brightest source in the two tiles
being on {\it B} and reaching into {\it A}. In this case, a fit for the
brightest object is not available for estimating the optimal sky
background for a number of galaxies on tile {\it A}. The underlying
idea of this method is that the average object size is much smaller
than the size of a survey image.

These two approaches a) and b) are implemented in the code as follows.
\sersic\ fitting with \gala\ is divided into two parts:

In the first part (see Fig.~\ref{fig_structure} upper section of block
D), \gala\ treats a fraction of all sources on all tiles in a sorted
order as laid out in method a). This assures that the brightest galaxy
from tile {\it B} is fitted before \gala\ treats tile {\it A} or {\it
B}. This part still requires sequential processing without the
possibility to run other instances of \gala\ at the same time. Also, it
produces a rather large computational overhead, as potentially with
every new source a number of large images (the complete science image,
weight image, segmentation map, etc. -- not the postage stamps) are to
be loaded into memory and processed (for fitting the sky background). A
possible working definition for the fraction of sources that have to be
fitted sequentially might encompass all sources that span an area
larger than the size of the overlaps resulting from the survey's tiling
scheme (\verb|D12|). This stage requires that all CPUs must be able to
see the whole dataset, \ie\ they have to have access to the same
harddiscs, because several threads are interacting with each other and
working on the same data.

The second part (see Fig.~\ref{fig_structure} lower section of block D)
is kept as detailed above in method b): \gala\ processes all objects
within a tile in order of decreasing brightness. Several instances of
\gala\ may be run simultaneously on {\it different} tiles. With the
sources that potentially reach into neighbouring tiles already
processed in the first part, now survey images may be treated as
individual entities, which can be processed out of order and
simultaneously. At this stage, one might think that, as the tiles are
decoupled, only a single tile is accessed at a time. However, in the
presence of big, bright sources that affect neighbouring tiles, this is
not the case any more. Therefore, even when fitting individual tiles in
parallel the whole data set must be accessible. As now only the
information for the current tile is changed, the fitting may be
distributed to different harddiscs, though (by creating identical
copies).

\begin{figure}\centering\resizebox{8cm}{!}{\includegraphics{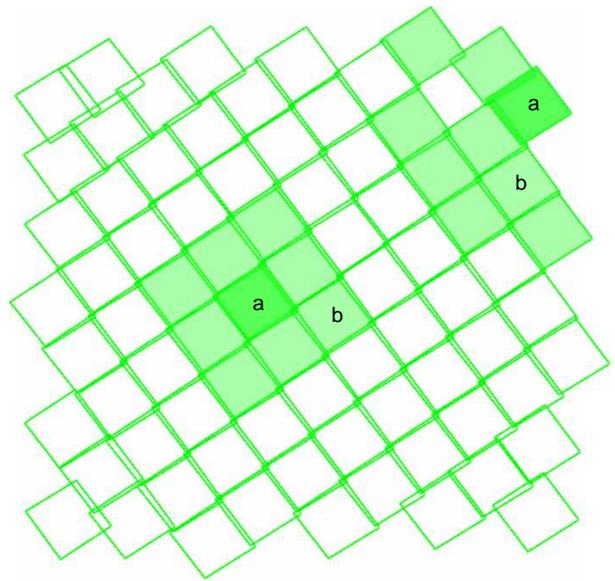}}
\caption{Definition of neighbouring tiles. For each survey image the
$n$ closest neighbours define its immediate neighbourhood. In a
checker-board configuration $n=8$ corresponds to a $3\times 3$ pattern.
At the edge of the survey more distant tiles are included. The light
green shaded areas {\it b} indicate the neighbourhoods for the central
tiles {\it a}.}\label{fig_neigh} \end{figure}

\subsubsection*{Neighbouring Tiles}

During the sky background estimation \gala\ calculates the influence of
all objects on the currently processed source. Depending on the size of
the survey, this check for contributors takes up a significant fraction
of the complete source loop computation time. However, the sources
immediately required for processing the current object are only the
ones that may have an impact on the fitting or background estimation.
Therefore, by specifying the ``reach'' of the brightest sources allows
to restrict the computation to a much smaller fraction of all sources.
This is done by providing the total number of closest tiles $n$ that
are to be included in the calculation (\verb|D18|). If the tiles are
taken on a regular grid, $n=8$ defines a ring surrounding the tile of
the current object (see Fig.~\ref{fig_neigh}). Note that in case of a
tile at the edge of the survey this ``ring'' is not cut in half, but
all tiles are selected on just one side. \gala\ {\it always} selects
the nearest $n$ neighbours. It calculates the distance between tiles
from the centres of the images.

\begin{figure}\centering\resizebox{8cm}{!}{\includegraphics{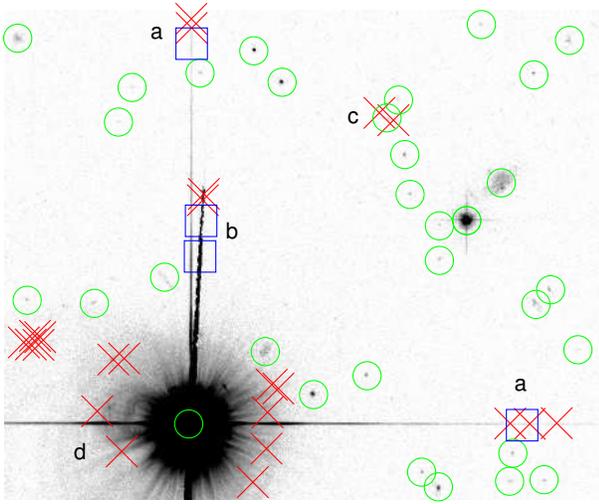}}
\caption{Correction of detection errors. Red crosses indicate
``critical'' detections; blue boxes mark ``catalogue'' detections;
green circles show ``good'' source detections. For a definition of
these terms see Sec.~\ref{sec_opt}. {\it a:} The diffraction spike of
the star was picked up as multiple individual sources. Most of them are
critical failures. Yet, one detection in each spike was kept as a
catalogue source, in order to guarantee that the fitting of nearby
objects is not biased. {\it b:} Pixel bleeding from the star. Again,
some detections were flagged critical, others as catalogue sources.
{\it c:} An over-deblended object. The excess detections are critical
errors. {\it d:} Spurious detections in the vicinity of the bright
star. All are critical failures. Note that the categorisation of
sources is an optional, subjective process, which is not performed
automatically by \gala.} \label{fig_sexerr} \end{figure}

\subsubsection*{Detection Flags}\label{sec_bad_detections}

A {\it perfect} setup for \sex\ never exists. In a small fraction of
all detections one or the other failure occurs, \eg\
(over-)~deblending, non-detection, spurious detection, etc. In
particular in the surroundings of bright stars (or even galaxies) these
errors accumulate. With respect to setting up \galfit\ properly, there
are two classes of failures: the ``critical'' and the ``catalogue''
failures. Depending on their relative brightness compared to nearby
``real'' objects they either have to be removed before the fitting
(faint sources; ``critical'') or after (bright sources; ``catalogue'').

A critical failure is a detection error that should be corrected {\it
before} running \galfit. Critical detections {\it do not affect the
fitting of neighbouring real sources}. Examples are over-deblends,
cosmic rays or a bad detection at the image edge. Critical failures
include any unwanted detection that might erroneously include
additional unnecessary components in the fitting of real objects. We
give an example for an over-deblended source in Fig.~\ref{fig_sexerr}
{\it c} and indicate several spurious detections in
Fig.~\ref{fig_sexerr} {\it d}.

In contrast, catalogue failures are detections that one has to remove
{\it after} running \galfit. They are bright in relation to
neighbouring sources and they {\it might affect the fitting of nearby
objects} if not included as separate components. Typically, they are
connected to cosmetic ``defects'' of the image. A common example for
these are diffraction spikes of stars, which may not be included in the
PSF model. Therefore, \galfit\ may not properly fit a galaxy close to
such a spike, as too much flux is in the spike compared to the source.
Common are also satellite trails or pixel column bleeding of saturated
stars. We show some examples in Fig.~\ref{fig_sexerr} {\it a} and {\it
b}.

\gala\ can optionally take care of both these failures. If the user
provides a (manually created) list of positions for critical and/or
catalogue failures (one file each), \gala\ will remove any source found
within a specified radius \verb|B16| around these positions from the
catalogues at the proper stage in the process. Otherwise, \gala\ treats
them simply as normal sources and provides \sersic\ fits for them. A
cleaner --although potentially somewhat more time consuming-- approach
would be to remove problematic regions from the data altogether (\eg\ by
replacing with white noise).

To classify unwanted detections (into one of the two categories), the
user should decide whether an object is required for obtaining a proper
fit with \galfit\ for neighbouring ``real'' sources, or not. In
principle it is save to put any detection error into the catalogue
failure list. This might lead to prolonged fitting times, though. In
practice, most detection errors are faint enough to not influence
neighbours and should therefore be put into the list of critical
failures.

Note that the definition of whether an object is a critical or
catalogue failure is subjective and depends on the user. However, the
correction of these errors is an option. \gala\ will run perfectly well
without any manual treatment. In that case, the user will have to live
with the fact that some (small) fraction of sources might be affected
by this.

\subsubsection*{Treatment of Stars}

A problem related to fitting bright saturated stars is that they are
often much brighter than the stars that one can use as a PSF model.
Because there is a limited dynamic range, the PSF cannot adequately
capture the tails seen around brighter stars, which may then
contaminate neighbouring galaxies. To deal with this situation, we fit
\sersic\ models to stars instead of the usual PSF model, because a high
\sersic\ index produces a model with extended tails. However, in so
doing, it may cause \galfit\ to not converge within a reasonable amount
of time. As the focus of \gala\ is on modelling the properties of
galaxies, no further attempt was made to apply a different, more
elaborate model (instead of the \sersic\ profile).

\begin{figure}\centering\resizebox{8cm}{!}{\includegraphics{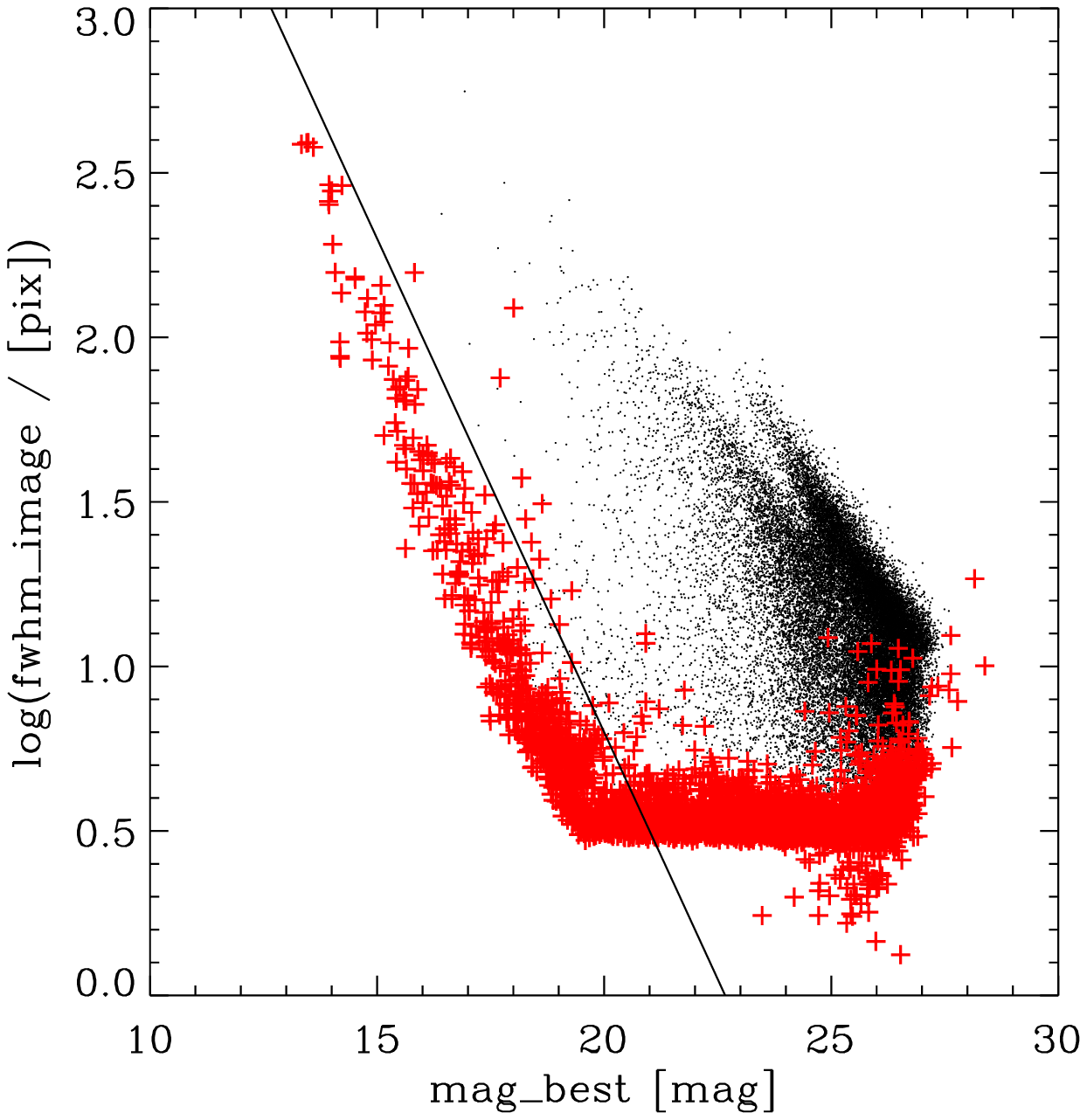}}
\caption{Treatment of saturated stars. Here we show source detections
from the STAGES survey in the $\log\left(\tt{fwhm\_image}\right)$
vs.~$\tt{mag\_best}$ plane. Red pluses mark objects identified as stars
\citep[see][]{ref_stages}. The line indicates the cut used to identify
saturated stars (left of the line). Black dots show other
(extragalactic) sources.}\label{fig_stars} \end{figure}

To resolve the resulting problem with the convergence of \galfit\,
\gala\ identifies saturated stars in the magnitude-size plane, which is
represented by the \sex\ parameters $\tt{mag\_best}$ and
$\log\left(\tt{fwhm\_image}\right)$ (see Fig.~\ref{fig_stars}). The
user specifies the zeropoint \verb|D15| and slope \verb|D14| of a line
below which \gala\ treats objects as saturated stars (\ie\ on the
bright and compact/small side). The reason for many of the brighter
stars to fail in the fitting is the detection of a large number of
neighbouring secondary sources (including stellar diffraction spikes),
which have to be modelled simultaneously. To reduce the number of these
secondaries the user specifies a relative magnitude cut \verb|D17|,
below which secondaries are not fitted any more and treated as
tertiaries. For the STAGES data, all objects more than two magnitudes
fainter than the primary star
($m_{\mathrm{star}}-m_{\mathrm{object}}>2$) were subject to this. Note
that for galaxies the same limitation applies, but at a much weaker
level. Again a magnitude limit \verb|D16| may be specified (\eg\
$m_{\mathrm{galaxy}}-m_{\mathrm{object}}>5$). Restricting the number of
secondaries to those objects bright enough to influence the fit and
removing the fainter ones resolves the issue.

\begin{figure*}\centering\resizebox{17.5cm}{!}{
\includegraphics{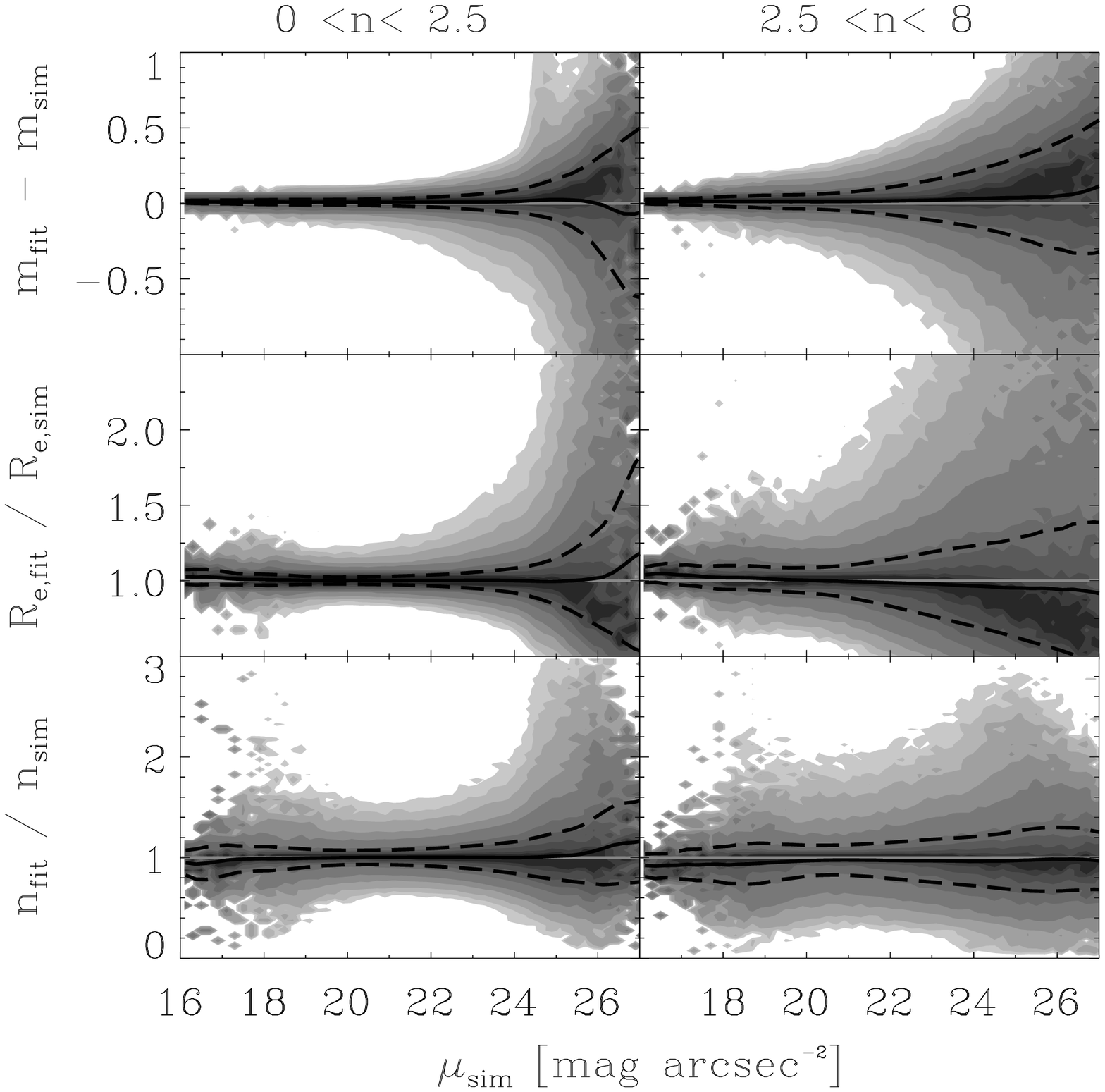}\hspace{0.5cm}
\includegraphics{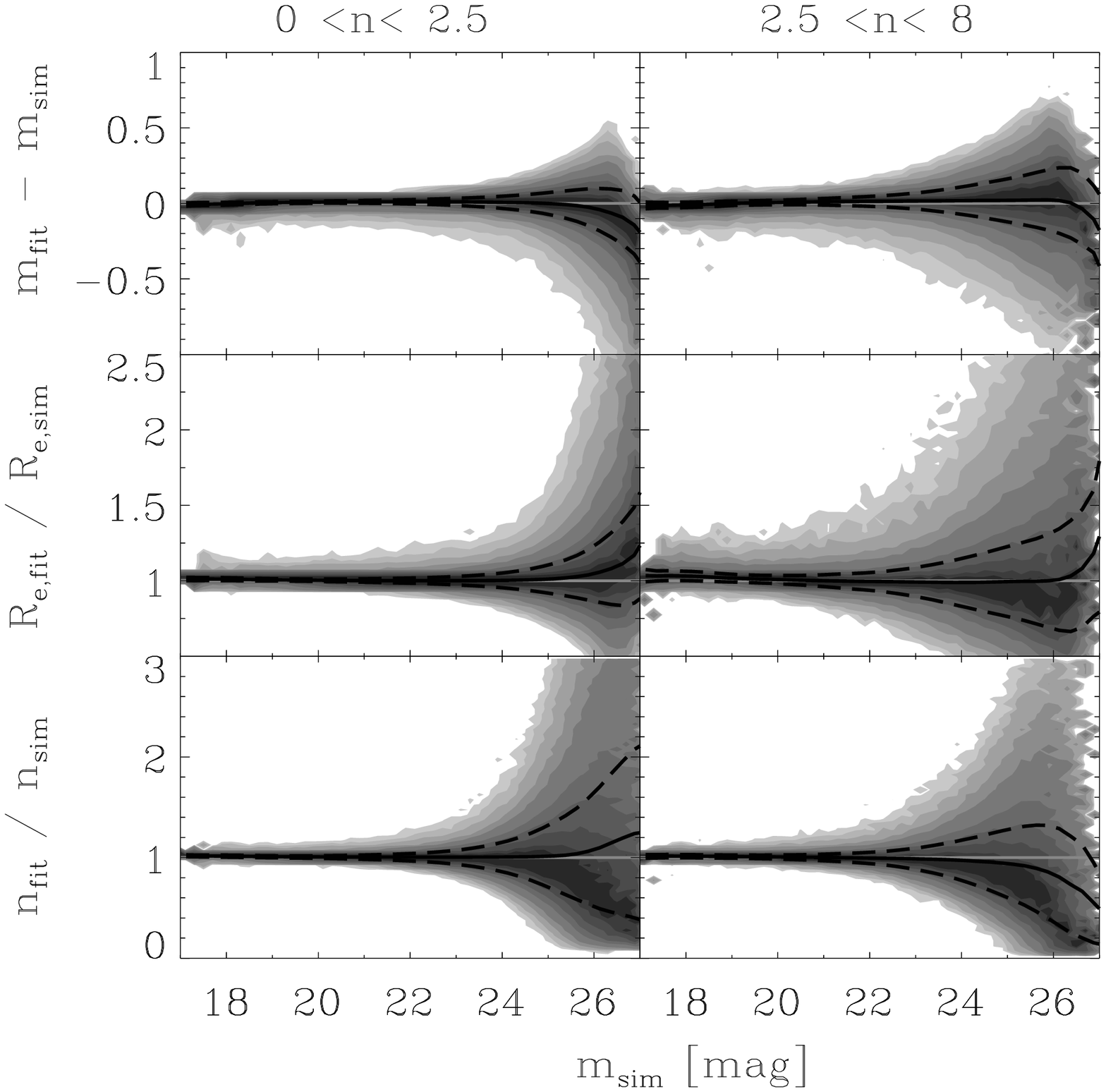}} \caption{Parameter recovery as a
function of simulated mean surface brightness $\mu_\mathrm{sim}$ within
$R_\mathrm{e}$ ({\it left panel}) and simulated magnitude
$m_\mathrm{sim}$ ({\it right panel}) for two different \sersic\ index
ranges (disc-like galaxies with $n\approx1$ on the left-hand side,
early-type galaxies with $n\approx4$ on the right-hand side). Grey
levels show galaxy density, with each bin being normalised to its own
peak value. As a result, grey levels roughly resemble a mean value and
a measure for the scatter of the distribution. Due to an asymmetric
distribution and different binning, the true mean value (black line)
deviates slightly from the peak values for fainter galaxies. The
1-$\sigma$ scatter of the distributions is shown as well (dashed
lines). The light grey line indicates the ideal zero-level. Fainter
galaxies (both as function of magnitude and surface brightness) and
galaxies with higher $n$ are fitted less accurately. Also, for the
brightest galaxies in the sample, the deviation increases. Most likely
their brightness (and size) makes them the most difficult objects to
setup for fitting, because of a large number of simultaneously fitted
neighbours and because of having the highest uncertainty in their
background sky estimate.}\label{fig_param_recovery} \end{figure*}

\begin{figure*} \centering\resizebox{17.5cm}{!}{
\includegraphics{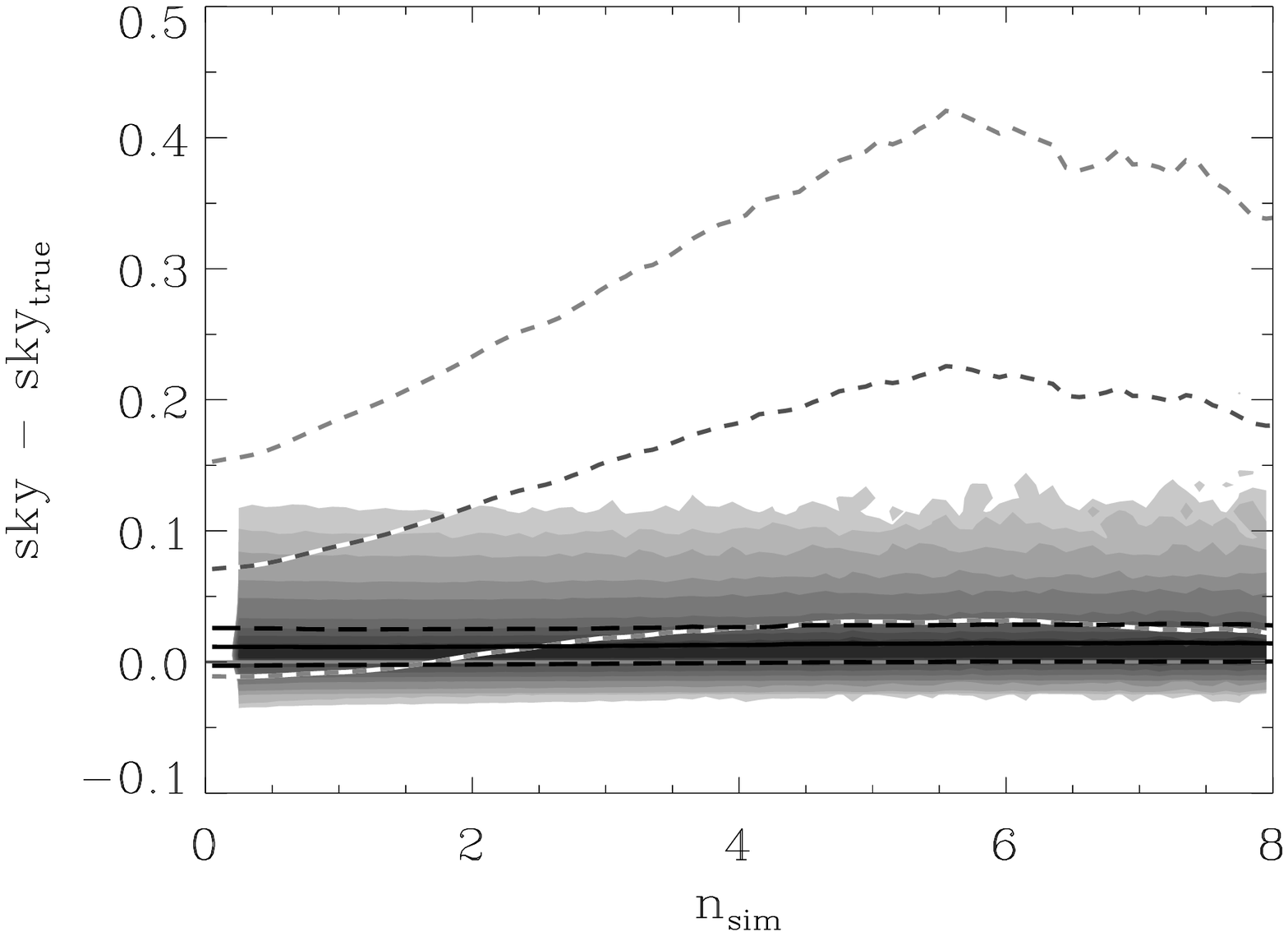}\hspace{0.5cm}
\includegraphics{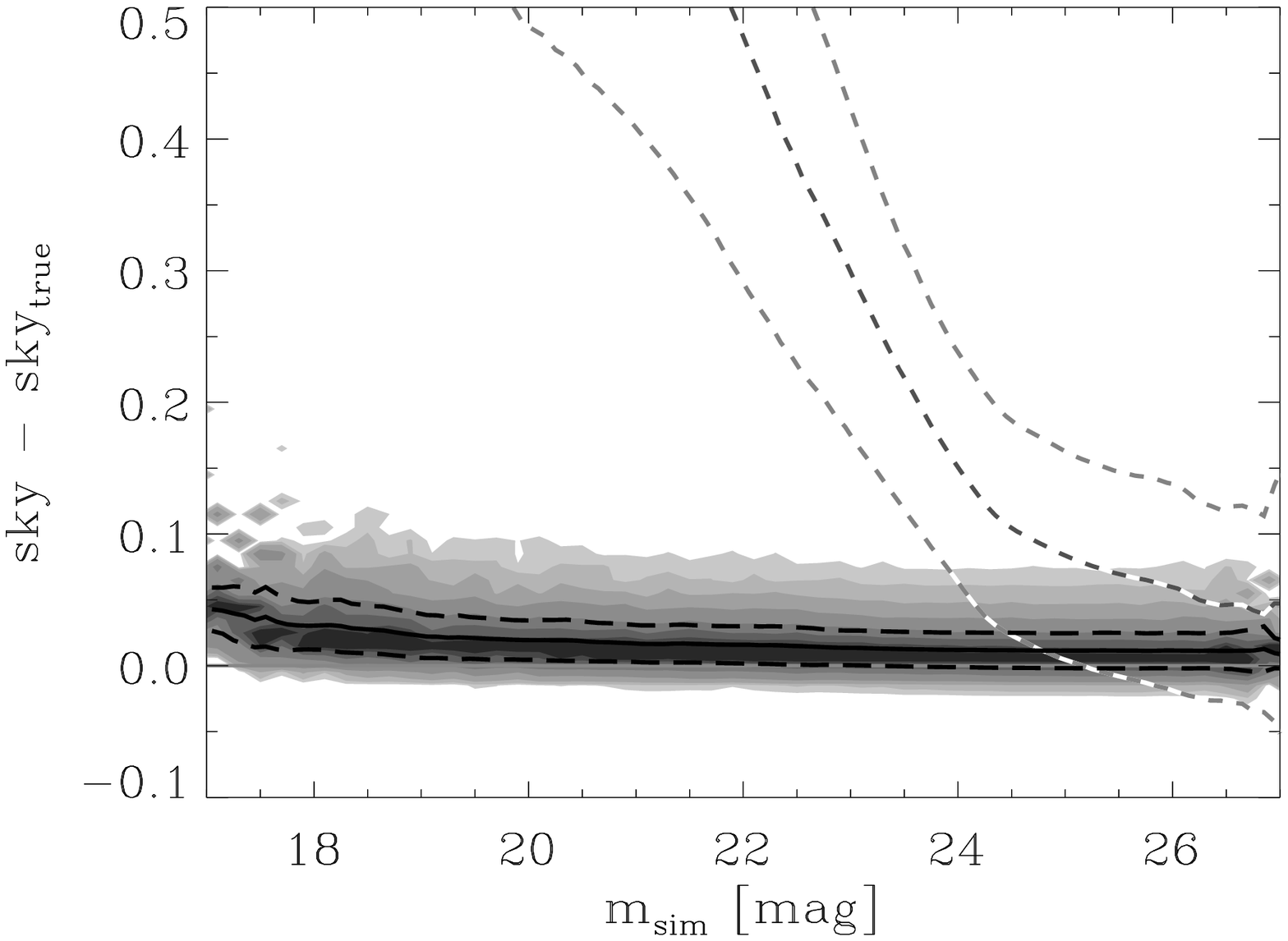}} \caption{Sky recovery (flux
difference in counts) as a function of simulated galaxy \sersic\ index
({\it left panel}) and simulated magnitude ({\it right panel}).
Contours and black lines show the distribution, mean and $\sigma$ of
the estimated sky as recovered by \gala, white/grey dashed lines
indicate mean and $\sigma$ as provided by \sex. \gala\ recovers a very
accurate sky value independent of galaxy structure, whereas \sex\
overall exhibits a much larger offset, scatter and dependence on galaxy
morphology. At the brightest magnitudes a slight trend is seen in
\gala\ while \sex\ performs worse by a factor of $\sim$50. Note that
the right panel shows only objects with $3<n<5$, and thus portrays a
rather conservative scenario.}\label{fig_sky_recovery} \end{figure*}

\section{Data Quality}\label{sec_quality}

We have tested \gala\ thoroughly using simulated data as described in
more detail in \cite{ref_fitting} and \cite{ref_stages}. For the
simulations applied here, we use the same setup as for fitting the
STAGES survey. Analytical \sersic\ profiles are randomly placed on a
background image composed from patches of blank sky from real data.
The galaxy models are convolved with the same PSF as the original
STAGES data (before placing them on the background). Also Poisson
noise was added to each pixel of the galaxy models. The galaxy model
parameters randomly cover the same parameter space as the original
STAGES data with an extension to towards low fluxes and surface
brightnesses, such as to cover the full completeness space.

All in all, the simulated datasets contain around 7
million galaxies. Excluding the ones that are not recovered by \gala\
for being below the detection threshold (3 million) and the ones that
ran into any given fitting constraint ($\sim$280\,000; the following
constraints for the \sersic\ index $n$, the half-light radius
$R_\mathrm{e}$ and the magnitude $m$ were applied: $0.2<n<8$,
$0.3<R_\mathrm{e}$~[pix] $<750$,
$|m_\mathrm{GALFIT}-m_\mathrm{SExtractor}|<5$) or where the fit crashed
(293), leaves us with around 3.7 million successfully fitted galaxies.

The left panel of Fig.~\ref{fig_param_recovery} shows the deviations of
the three most important fitting parameters magnitude $m$, effective
radius $R_\mathrm{e}$ and \sersic\ index $n$ as a function of simulated
mean surface brightness $\mu_\mathrm{sim}$ within $R_\mathrm{e}$ for
two different regimes of \sersic\ index. We choose the samples such
that the completeness as a function of magnitude is roughly 90\% for
all galaxies. The low \sersic\ index sample ($m_\mathrm{sim}<24.5$)
contains $\sim$1.1 million galaxies, the high \sersic\ sample
($m_\mathrm{sim}<25.25$) contains $\sim$470\,000 galaxies. Obviously,
\gala' performance decreases at faint magnitudes and high \sersic\
indices. The right panel of Fig.~\ref{fig_param_recovery} shows the
same plot, but as a function of simulated magnitude $m_\mathrm{sim}$
rather than surface brightness to illustrate the same effects in
another commonly used parameter space. Again, we choose a cut to select
only galaxies with a surface brightness completeness exceeding 90\%.
The low \sersic\ index sample ($\mu_\mathrm{sim}<22.25$) contains
$\sim$780\,000 galaxies, the high \sersic\ sample
($\mu_\mathrm{sim}<23$) contains $\sim$295\,000 galaxies. At the faint
end, quite expectedly, the recoverability of parameters gets worse. In
both panels of Fig.~\ref{fig_param_recovery}, we see no significant
systematic trends apart from the faintest levels.

The left panel in Fig.~\ref{fig_sky_recovery} shows the deviation of
the sky value (as recovered by \gala) from the true sky value (as
derived from the empty noise image used for the galaxy simulations) as
a function of the simulated \sersic\ index $n$ of the primary object.
Obviously, the recovery of the sky in \gala\ is completely independent
of $n$. Compared to the \sex\ value for the local sky, which shows both
a much bigger offset and a larger standard deviation, the recovery is
close to ideal with very small offset and scatter. We derive the true
sky value for this plot from simple statistics on an empty noise image
used for the simulations.

\begin{figure}
\centering\resizebox{8cm}{!}{\includegraphics{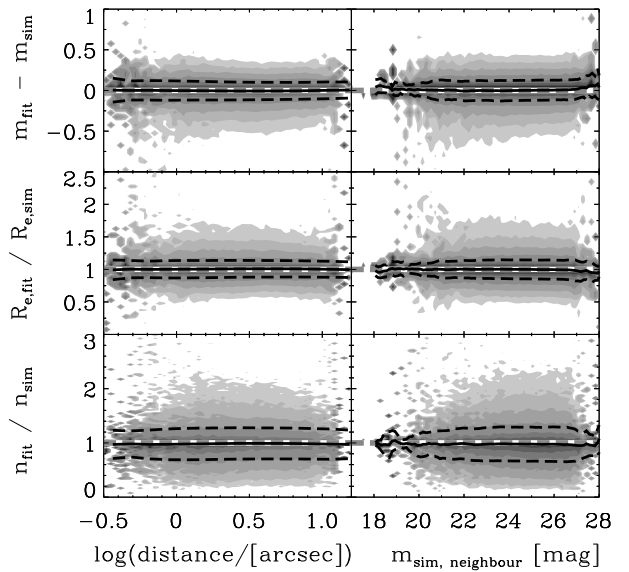}}
\caption{Parameter deviations as a function of both distance ({\it left
panel}) and magnitude ({\it right panel}) of the nearest neighbouring
galaxy. The thick grey/white dashed lines indicate the zero-level;
black solid and dashed lines show the mean and $\sigma$ of the
recovered parameter, respectively. Grey contours represent the
normalised distribution of recovered parameter
values.}\label{fig_neighbour_influence} \end{figure}

Furthermore, we investigate the magnitude dependence of the sky
recoverability (see right panel of Fig.~\ref{fig_sky_recovery}). Here
we select only objects with a \sersic\ index $3<n<5$ ($\sim$300\,000
objects), which due to their extended low surface brightness wings are
hardest to fit and estimate a background value. Thus, we portray a
conservative worst case scenario. While for the large majority of all
objects there is no trend to be seen at all, at the bright end the
estimates provided by \gala\ do diverge slightly: at $m=17,18$ the mean
sky moves off by $\sim$0.04/0.03 with a scatter of $\sim$0.02,
respectively. For comparison, the values recovered by \sex\ are
$\sim$2.3/1.3 with a scatter of $\sim$0.4 at the same brightness.

To examine the influence of neighbouring galaxies in a similar fashion
as was shown in \cite{ref_fitting}, we plot parameter deviations over
both magnitude of and distance from the next neighbour. The next
neighbour we here define as the closest simulated galaxy that was found
by \sex. This does not necessarily imply that this galaxy had to be
properly deblended and simultaneously fitted when running \galfit\
(\ie\ assuming a rather conservative definition resulting in a worst
case scenario). We show these deviations in
Fig.~\ref{fig_neighbour_influence}. In contrast to the analysis in
\cite{ref_fitting}, we now have enough statistical significance to
separate both effects. We only show neighbours with
$21<m_\mathrm{sim}<23$ ($\sim$330\,000 galaxies) in the left panel and
neighbours with a distance $1<d$ [arcsec] $<1.6$ ($\sim$280\,000
galaxies) in the right panel of Fig.~\ref{fig_neighbour_influence} to
not confuse the two distinct effects: contamination by bright
neighbours and contamination by close neighbours. As one can see from
these plots, \gala\ results do not show any dependence on either of
these parameters. From this plot we conclude that the deblending and
fitting scheme applied in \gala\ works well and successful deblending
of clustered fields (as \eg\ STAGES) is possible.

\begin{figure}\centering\resizebox{8cm}{!}{\includegraphics{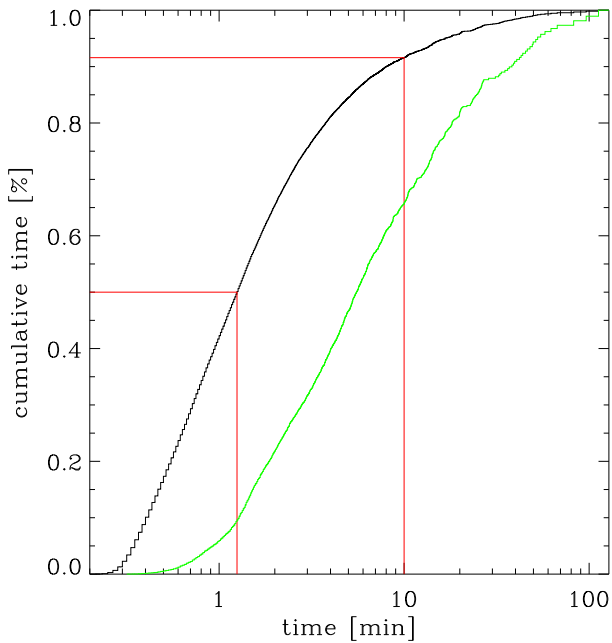}}
\caption{Performance of the galaxy modelling with \galfit. Cumulative
histogram of the fitting time per object as a fraction of the total
fitting time. The two histograms show all galaxies (black/left) and the
brightest 5\% (green/right).50\% of all sources take less than 1.25~min
to fit and more than 90\% of all objects are done within 10~min (red
lines). The brightest 5\% take about a factor of 5
longer.}\label{fig_performance} \end{figure}

\section{Performance}\label{sec_performance}

We measure the performance of \gala\ when applying it to the
single-orbit \hst\ survey STAGES \citep[see][]{ref_stages}. STAGES is a
mosaic composed of 80 tiles in the F606W filter containing
$\sim$75\,000 sources. The survey being centred on a nearby galaxy
cluster system at redshift $z\sim$0.16, it provides the ideal test case
including a high fraction of large and also peculiar objects. Large
objects serve as a test for the deblending process during the source
extraction, while peculiar objects like mergers or saturated stars with
diffraction spikes pose a challenge for the modelling with \galfit. The
total wall-clock time for processing this survey is $\sim$430~hours.
Details are given below.

The fitting process with \galfit\ is the main limitation for \gala\ '
performance. The largest amount of time was spent on fitting the
fainter 95\% of all sources in the parallel mode (second part of block
\verb|(D)| in Fig.~\ref{fig_structure}). Using eight 2.2~GHz CPU cores
in parallel, this process (\ie\ the slowest of the eight) takes
$\sim$260~hours. There is potential for further improvement by
increasing the total number of CPU cores. This would also reduce the
overhead resulting from individual pipelines not finishing at the same
time (\ie\ pipelines with fewer sources finish sooner), resulting
normally in much less than the total number of available CPUs running
simultaneously at the end of the fitting.

For the first part of block \verb|(D)|, the fitting of the brightest
5\% of all sources we used four 2.4~GHz CPU cores. This part of the fitting
takes $\sim$150~hours. Note that moving from four CPU cores to eight does
not necessarily imply halving the required computation time. The
performance increase at this stage depends on the survey geometry. A
wide area survey has a higher efficiency than a smaller survey of the
same depth, because of the higher probability that the brightest
objects in the survey are further apart from each other, thus allowing
a higher multiplicity. Fig.~\ref{fig_performance} shows a cumulative
histogram of the fitting time per object. Note that the brightest
objects take considerably longer to fit than the rest thus explaining
the necessity to find a good compromise between the time spent in the
two stages.

The remaining blocks take up an almost negligible fraction of the total
processing time. Block \verb|(B)|, the \sex\ stage, takes
$\sim$13.5~hours, including HDR mode. Cutting the postage stamps in
block \verb|(C)| requires $\sim$2.5~hours and the last block
\verb|(F)|, compilation of the output catalogue, finishes within
$\sim$0.7~hours.

Note that overheads for adjusting the setup and preparing the parallel
fitting is not taken into account in the numbers cited above. Also, for
varying survey layouts/configurations relative fractions of the total
processing times between the various stages might vary significantly.

\section{Summary}\label{sec_summary}

We present \gala, a software for automating the process of detecting
sources and modelling them with single \sersic\ profiles. \gala\
incorporates \sex\ and \galfit\ to perform these two tasks. In
addition, it provides HDR source extraction, a postage stamp cutting
facility and a robust means of estimating a local sky background. It
stores results in a combined FITS table, excluding duplicates resulting
from detections in overlapping tiles. We optimised the code for speed
and stability, making use of modern multi-core CPUs and allowing a high
degree of multiplicity. Another aim was to present the user with a
simple setup, yet enabling control over all features of the code. As a
result, \gala\ can be used on a wide variety of survey applications,
from single tile deep observations to wide area shallow surveys.
\galfit 's ability to work with any given PSF enables application of
\gala\ to both space- and ground-based data. The PSF has to be prepared
by the user before running \gala, though. This procedure is not part of
the code.

We tested \gala\ on an
extensive set of simulations and find it to be extremely robust in
terms of parameter recoverability. Note that the results of the fitting
depend on the choice of the input parameters. For example, a bad \sex\
setup will have a significant impact on the fitting procedure and thus
lower the quality of the output catalogue.

The main feature that will be implemented in \gala\ in the near future
is the option for a consistent two-component bulge-disc fitting. This
will also include an estimator providing information about whether the
increased amount of data potentially allows further insight into the
structural composition of the object or not. Based on this idea, we
will also investigate the automated fitting of bars and the application
of Fourier mode fitting, built into the most recent version of \galfit.

Another potential aspect for increasing the versatility of \gala\ could
be the implementation of a variable PSF. Currently, just one PSF is
used for convolving the \galfit\ model profiles for the whole survey.
Instead, one could allow using a different PSF depending on the
position on the tile or even varying tile by tile.

\gala\ is freely available for download from our webpage at:
\url{http://astro.uibk.ac.at/~m.barden/GALAPAGOS/}

\section*{acknowledgements}

MB was supported in part by the \emph{Austrian Science Foundation FWF} under
grant P18416. BH is grateful for support from the \emph{Science and Technology
Facilities Council (STFC)}. DHM acknowledges support from the \emph{National
Aeronautics and Space Administration (NASA)} under LTSA Grant NAG5-13102 issued
through the \emph{Office of Space Science}. CYP acknowledges support from the
Canadian NRC-HIA Plaskett Fellowship and the STScI Institute/Giacconi Fellowship
programs.

\bibliographystyle{mn2e}
\bibliography{mn-jour,galapagos}

\begin{thebibliography}{21}
\expandafter\ifx\csname natexlab\endcsname\relax\def\natexlab#1{#1}\fi

\bibitem[{{Beckwith} {et~al}\mbox{.}(2006){Beckwith}, {Stiavelli}, {Koekemoer},
  {Caldwell}, {Ferguson}, {Hook}, {Lucas}, {Bergeron}, {Corbin}, {Jogee},
  {Panagia}, {Robberto}, {Royle}, {Somerville}, \& {Sosey}}]{ref_hudf}
{Beckwith} S.~V.~W. {et~al.}, 2006, AJ, 132, 1729

\bibitem[{{Bertin} \& {Arnouts}(1996)}]{ref_sex}
{Bertin} E., {Arnouts} S., 1996, A\&AS, 117, 393

\bibitem[{{Caldwell} {et~al}\mbox{.}(2008){Caldwell}, {McIntosh}, {Rix},
  {Barden}, {Beckwith}, {Bell}, {Borch}, {Heymans}, {H{\"a}u{\ss}ler},
  {Jahnke}, {Jogee}, {Meisenheimer}, {Peng}, {S{\'a}nchez}, {Somerville},
  {Wisotzki}, \& {Wolf}}]{ref_gems_cat}
{Caldwell} J.~A.~R. {et~al.}, 2008, ApJS, 174, 136

\bibitem[{{de Jong}(1996)}]{ref_dejong}
{de Jong} R.~S., 1996, A\&AS, 118, 557

\bibitem[{{de Souza}, {Gadotti} \& {dos Anjos}(2004){de Souza}, {Gadotti}, \&
  {dos Anjos}}]{ref_budda}
{de Souza} R.~E., {Gadotti} D.~A., {dos Anjos} S., 2004, ApJS, 153, 411

\bibitem[{{Giavalisco} {et~al}\mbox{.}(2004){Giavalisco}, {Ferguson},
  {Koekemoer}, {Dickinson}, {Alexander}, {Bauer}, {Bergeron}, {Biagetti},
  {Brandt}, {Casertano}, {Cesarsky}, {Chatzichristou}, {Conselice},
  {Cristiani}, {Da Costa}, {Dahlen}, {de Mello}, {Eisenhardt}, {Erben}, {Fall},
  {Fassnacht}, {Fosbury}, {Fruchter}, {Gardner}, {Grogin}, {Hook},
  {Hornschemeier}, {Idzi}, {Jogee}, {Kretchmer}, {Laidler}, {Lee}, {Livio},
  {Lucas}, {Madau}, {Mobasher}, {Moustakas}, {Nonino}, {Padovani}, {Papovich},
  {Park}, {Ravindranath}, {Renzini}, {Richardson}, {Riess}, {Rosati},
  {Schirmer}, {Schreier}, {Somerville}, {Spinrad}, {Stern}, {Stiavelli},
  {Strolger}, {Urry}, {Vandame}, {Williams}, \& {Wolf}}]{ref_goods}
{Giavalisco} M. {et~al.}, 2004, ApJ, 600, L93

\bibitem[{{Gray} {et~al}\mbox{.}(2008){Gray}, {Wolf}, {Barden}, {Peng},
  {H{\"a}u{\ss}ler}, {Bell}, {McIntosh}, {Caldwell}, {Bacon}, {Balogh},
  {Barazza}, {B{\"o}hm}, {Heymans}, {Jahnke}, {Jogee}, {van Kampen}, {Lane},
  {Meisenheimer}, {S{\'a}nchez}, {Taylor}, {Wisotzki}, {Zheng}, {Green},
  {Beswick}, {Saikia}, {Gilmour}, {Johnson}, \& {Papovich}}]{ref_stages}
{Gray} M.~E. {et~al.}, 2008, ArXiv e-prints

\bibitem[{{H{\"a}ussler} {et~al}\mbox{.}(2007){H{\"a}ussler}, {McIntosh},
  {Barden}, {Bell}, {Rix}, {Borch}, {Beckwith}, {Caldwell}, {Heymans},
  {Jahnke}, {Jogee}, {Koposov}, {Meisenheimer}, {S{\'a}nchez}, {Somerville},
  {Wisotzki}, \& {Wolf}}]{ref_fitting}
{H{\"a}ussler} B. {et~al.}, 2007, ApJS, 172, 615

\bibitem[{{Infante}(1987)}]{ref_infante}
{Infante} L., 1987, A\&A, 183, 177

\bibitem[{{Koekemoer} {et~al}\mbox{.}(2007){Koekemoer}, {Aussel}, {Calzetti},
  {Capak}, {Giavalisco}, {Kneib}, {Leauthaud}, {Le F{\`e}vre}, {McCracken},
  {Massey}, {Mobasher}, {Rhodes}, {Scoville}, \& {Shopbell}}]{ref_cosmos_hst}
{Koekemoer} A.~M. {et~al.}, 2007, ApJS, 172, 196

\bibitem[{{Kron}(1980)}]{ref_kron}
{Kron} R.~G., 1980, ApJS, 43, 305

\bibitem[{{Lauer} {et~al}\mbox{.}(1995){Lauer}, {Ajhar}, {Byun}, {Dressler},
  {Faber}, {Grillmair}, {Kormendy}, {Richstone}, \& {Tremaine}}]{ref_nuker}
{Lauer} T.~R. {et~al.}, 1995, AJ, 110, 2622

\bibitem[{{Leauthaud} {et~al}\mbox{.}(2007){Leauthaud}, {Massey}, {Kneib},
  {Rhodes}, {Johnston}, {Capak}, {Heymans}, {Ellis}, {Koekemoer}, {Le
  F{\`e}vre}, {Mellier}, {R{\'e}fr{\'e}gier}, {Robin}, {Scoville}, {Tasca},
  {Taylor}, \& {Van Waerbeke}}]{ref_cosmos_lensing}
{Leauthaud} A. {et~al.}, 2007, ApJS, 172, 219

\bibitem[{{Peng} {et~al}\mbox{.}(2002){Peng}, {Ho}, {Impey}, \&
  {Rix}}]{ref_galfit}
{Peng} C.~Y., {Ho} L.~C., {Impey} C.~D., {Rix} H.-W., 2002, AJ, 124, 266

\bibitem[{{Peng} {et~al}\mbox{.}(2010){Peng}, {Ho}, {Impey}, \&
  {Rix}}]{ref_galfit3}
---, 2010, AJ, 139, 2097

\bibitem[{{Rix} {et~al}\mbox{.}(2004){Rix}, {Barden}, {Beckwith}, {Bell},
  {Borch}, {Caldwell}, {H{\"a}ussler}, {Jahnke}, {Jogee}, {McIntosh},
  {Meisenheimer}, {Peng}, {Sanchez}, {Somerville}, {Wisotzki}, \&
  {Wolf}}]{ref_gems}
{Rix} H.-W. {et~al.}, 2004, ApJS, 152, 163

\bibitem[{{Scoville} {et~al}\mbox{.}(2007){Scoville}, {Aussel}, {Brusa},
  {Capak}, {Carollo}, {Elvis}, {Giavalisco}, {Guzzo}, {Hasinger}, {Impey},
  {Kneib}, {LeFevre}, {Lilly}, {Mobasher}, {Renzini}, {Rich}, {Sanders},
  {Schinnerer}, {Schminovich}, {Shopbell}, {Taniguchi}, \&
  {Tyson}}]{ref_cosmos}
{Scoville} N. {et~al.}, 2007, ApJS, 172, 1

\bibitem[{{Sersic}(1968)}]{ref_sersic}
{Sersic} J.~L., 1968, {Atlas de galaxias australes}. Cordoba, Argentina:
  Observatorio Astronomico, 1968

\bibitem[{{Simard} {et~al}\mbox{.}(2002){Simard}, {Willmer}, {Vogt},
  {Sarajedini}, {Phillips}, {Weiner}, {Koo}, {Im}, {Illingworth}, \&
  {Faber}}]{ref_gim2d}
{Simard} L. {et~al.}, 2002, ApJS, 142, 1

\bibitem[{{Vogt} {et~al}\mbox{.}(2005){Vogt}, {Koo}, {Phillips}, {Wu}, {Faber},
  {Willmer}, {Simard}, {Weiner}, {Illingworth}, {Gebhardt}, {Gronwall},
  {Guzm{\'a}n}, {Im}, {Sarajedini}, {Groth}, {Rhodes}, {Brunner}, {Connolly},
  {Szalay}, {Kron}, \& {Blandford}}]{ref_groth_strip}
{Vogt} N.~P. {et~al.}, 2005, ApJS, 159, 41

\bibitem[{{Wolf} {et~al}\mbox{.}(2004){Wolf}, {Meisenheimer}, {Kleinheinrich},
  {Borch}, {Dye}, {Gray}, {Wisotzki}, {Bell}, {Rix}, {Cimatti}, {Hasinger}, \&
  {Szokoly}}]{ref_combo}
{Wolf} C. {et~al.}, 2004, A\&A, 421, 913

\end{thebibliography}

\appendix

\section{Code Setup and Control}\label{sec_setup}

\begin{figure*}
\centering\includegraphics[height=23cm]{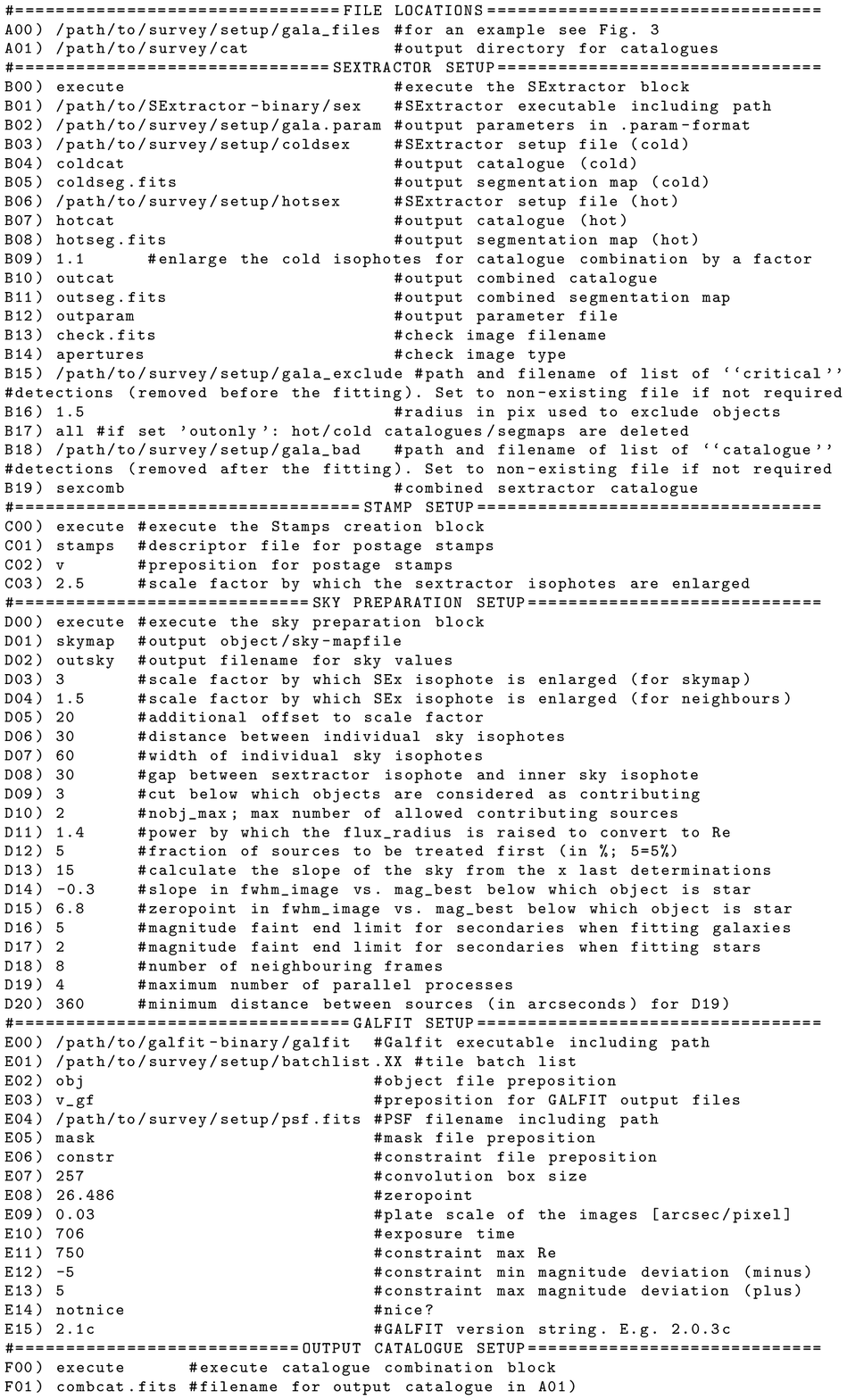}
\caption{Example of a startup script for \gala.}\label{fig_setupfile}
\end{figure*}

\begin{figure*}
%\centering\includegraphics[width=12cm]{gala_filelist.eps}
\centering\includegraphics{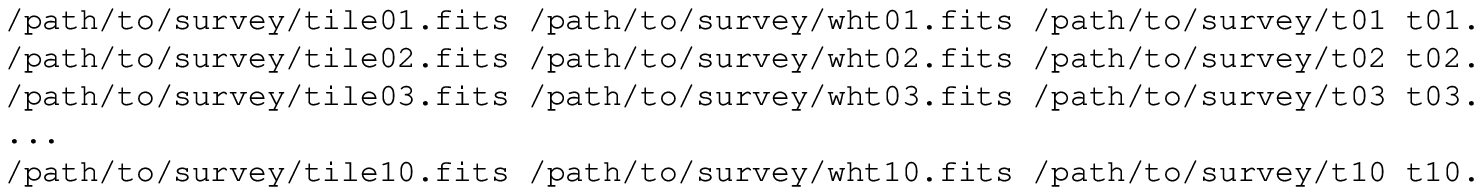} \caption{Example of a
file location list for \gala. The middle section is left out. In this
example a survey with 10 tiles is defined. The four columns represent
science image, corresponding weight image (exposure time map), output
directory and output file preposition,
respectively.}\label{fig_filelist}
\end{figure*}

In the following section we provide detailed information about how
\gala\ is run. This includes a description of the structure of setup
files and the execution sequence.

\subsection{The Setup Script}\label{sec_setup_main}

\gala\ is controlled by a set of scripts. We show an example for the
main startup script in Fig.~\ref{fig_setupfile}. It contains all
references for file locations and manages the programme execution. The
startup script is divided into six parts, \verb|(A)| through \verb|(F)|,
closely related to the four blocks described in
Sec.~\ref{sec_structure}. The first set of parameters \verb|(A)|
defines the input and output file locations. Section \verb|(E)|
contains options that help setting up \galfit. These two parameter sets
are ``static''; the remaining four are ``dynamic'' in the sense that
these can be activated or skipped when executing \gala. They correspond
to the four programme blocks that were previously defined in
Sec.~\ref{sec_structure}. Parameter set \verb|(B)| starts \sex;
\verb|(C)| is responsible for defining and cutting the postage stamps;
\verb|(D)| performs the estimation of the sky background, prepares
\galfit\ and starts the fitting; finally parameter set \verb|(F)| reads
out the fit results and creates the output catalogue. The difference
between the ``static'' and ``dynamic'' parameter sets is, that the
latter control code execution while the prior only define file
locations and setup parameters. Note that the setup files for \sex\
\verb|B03| (and optionally \verb|B06| in HDR-mode) require additional
files to be accessible. These are the neural network file
``default.nnw'' and an optional convolution filter, \eg\
``tophat\_3.0\_3x3.conv''. For details on how to setup \sex\ see
\cite{ref_sex}.

\subsection{The File List}

A file location list, or short ``file list'' \verb|A00|, provides the
information required for defining the organisation of the survey. It is
an ASCII file providing for each survey pointing it the file location
of the actual tile, the corresponding weight image (which is an
exposure time map), a path for storing the fitting output of the
individual tiles, and a prefix, which is attached to all output files.
We give an example of such a file list in Fig.~\ref{fig_filelist} for a
hypothetical survey with 10 science tiles.

\subsection{Catalogue Finetuning}

In order to refine the output of \sex, the user has the option to
remove sources. After an initial run with the optimal \sex\ setup, the
user may create a list that contains the file name of the respective
tile together with an x/y-pixel position, \eg:

\begin{verbatim}
/path/to/survey/tile01.fits 234 567
/path/to/survey/tile01.fits 765 432
/path/to/survey/tile02.fits 453 678
...
\end{verbatim}

\gala\ rejects any detection within a certain radius \verb|B16|
automatically from the \sex\ catalogue on a subsequent run of the code.
Thus, if one wants to refine the catalogue, the \sex\ section of the
code has to be run twice, \ie\ \gala\ needs to be started first with
only the \sex\ section activated and then run a second time with the
\sex\ section and optionally others enabled as well. The first
execution is required for identifying bad detections; the second run
then treats them. For details on what sources should be removed and how
the code deals with them see Sec.~\ref{sec_bad_detections}.

\subsection{Batch Processing}\label{sec_setup_batchlist}

In Sec.~\ref{sec_opt} we explain in some depth the mechanisms to
optimise the total programme execution time. According to this scheme,
after cutting the postage stamps and fitting a subset of all sources,
namely the brightest objects, \gala\ may be run in parallel on several
computers, each working on a section of the survey. In order to specify
the region that the respective pipeline should work on, the user must
provide a list \verb|E01| that contains the file names of the
individual tiles in question. If one were to fit the survey from
Fig.~\ref{fig_filelist} in parallel on three CPU cores, one could set up
the
first computer with a batch list containing tiles 1 to 3, the second
one with 4 to 6 and the third one with 7 to 10. As an example, the
batch list for the first CPU core would look like this:

\begin{verbatim}
/path/to/survey/tile01.fits
/path/to/survey/tile02.fits
/path/to/survey/tile03.fits
\end{verbatim}

\subsection{An Example Sequence}

To summarise, the process to run \gala\ on a complete survey requires
the
following steps:
\begin{enumerate}
 \item setup startup script \& file list (incl. \sex)
 \item run first block \verb|(B)| (optionally in HDR mode)
 \item optionally identify ``bad'' detections
 \item manually create the respective ``bad detection lists''
 \item if ``bad detection lists'' were created re-run block
       \verb|(B)|
 \item run block \verb|(C)| to prepare \& cut postage stamps
 \item run block \verb|(D)| on brightest galaxies
 \item create batch lists for parallel processing
 \item create startup scripts for batch lists
 \item re-run block \verb|(D)| in parallel on several machines
 \item when parallel processing is finished, run block
       \verb|(F)|
\end{enumerate}

Note that if the survey is small enough, steps 6~to 8~may be combined,
either by setting the brightest galaxies fraction \verb|D12| to 100\%,
thus taking full advantage of the available CPU cores on the machine or
by
simply providing only one batch file containing all tiles. The latter
option does not provide advantages over the first one and is best used
only for testing purposes.

\section{Starting Parameters}

In this appendix we give a detailed description of the starting
parameters in the \gala\ startup file. Lines starting with ``\#'' are
treated as comments and are ignored by the code. Examples for \sex\
related setup files (items \verb|B02|, \verb|B03|, \verb|B06| in the
table below) can be found in the corresponding documentation
\citep[][]{ref_sex}. In order not to execute the blocks \verb|B00|,
\verb|C00|, \verb|D00| or \verb|F00| the user should either replace
``execute'' with something else or simply comment out the respective
line, \eg\ ``\#B00) execute''. Files in the table below without a
directory descriptor can be found in the output directory defined for
each survey image in the file list unless otherwise noted.

\clearpage
\onecolumn

%\caption{Nonlinear Model Results}              % title of Table
%\label{table:1}      % is used to refer this table in the text
\centering
\begin{supertabular}{p{1cm} p{5.5cm} p{9cm}}
\hline\hline
\multicolumn{3}{c}{File Locations}\\
\hline
A00) & /path/to/survey/setup/gala\_files   & setup of the survey tiling (path
and filename). For an example see Fig.~\ref{fig_filelist}\\
A01) & /path/to/survey/cat                 & output directory for catalogues.
In this directory, the combined \sex\ catalogue (item B19); in ASCII
format) and the final output catalogue (item F01); in FITS format) are
placed\\
\hline\hline
\multicolumn{3}{c}{{\sc SExtractor} Setup}\\
\hline
B00) & execute                             & execute the \sex\ block\\
B01) & /path/to/SExtractor-binary/sex      & path and filename of \sex\
executable\\
B02) & /path/to/survey/setup/gala.param    & path and filename of \sex\ output
parameters in .param-format\\
B03) & /path/to/survey/setup/coldsex       & path and filename of \sex\ setup
file (cold)\\
B04) & coldcat                             & filename of \sex\ output catalogue
(cold)\\
B05) & coldseg.fits                        & filename of \sex\ output
segmentation map (cold)\\
B06) & /path/to/survey/setup/hotsex        & path and filename of \sex\ setup
file (hot)\\
B07) & hotcat                              & filename of \sex\ output catalogue
(hot)\\
B08) & hotseg.fits                         & filename of \sex\ output
segmentation map (hot)\\
B09) & 1.1                                 & factor by which the cold isophotes
are enlarged when combining hot/cold catalogues\\
B10) & outcat                              & filename of combined \sex\ output
catalogue\\
B11) & outseg.fits                         & filename of combined \sex\ output
segmentation map\\
B12) & outparam                            & filename of \sex\ output parameter
file\\
B13) & check.fits                          & filename of \sex\ check image\\
B14) & apertures                           & type of \sex\ check image\\
B15) & /path/to/survey/setup/gala\_exclude & path and filename of list of
``critical'' detections (removed {\it before} the fitting). Set to non-existing
file if not required\\
B16) & 1.5                                 & radius in pix used to exclude
``bad'' detections\\
B17) & all                                 & if set ``outonly'': hot/cold
catalogues/segmaps are deleted, else all files are kept\\
B18) & /path/to/survey/setup/gala\_bad     & path and filename of list of
``catalogue'' detections (removed {\it after} the fitting). Set to non-existing
file if not required\\
B19) & sexcomb                             & filename of combined \sex\
catalogue. Output directory is A01)\\
\hline\hline
\multicolumn{3}{c}{Stamp Setup}\\
\hline
C00) & execute & execute the postage stamps creation block\\
C01) & stamps & output descriptor file for postage stamps. Per line,
this ASCII
file contains: \sex\ number, x/y source centre, x-range, y-range\\
C02) & v       & filename preposition for postage stamps. E.g.~for C02) =
``v'', a global file preposition ``im1.'' (from file list) and \sex\
detection number ``234'', the output filename would be: ``im1.v234.fits''
\\
C03) & 2.5     & scale factor by which the \sex\ isophotes (Kron ellipses) are
enlarged to calculate postage stamp size\\
\hline\hline
\multicolumn{3}{c}{Sky Preparation}\\
\hline
D00) & execute & execute the sky preparation block\\
D01) & skymap  & filename of output object/sky-mapfile\\
D02) & outsky  & filename of output list with sky values\\
D03) & 3       & scale factor by which \sex\ isophote is enlarged (for
calculating the skymap)\\
D04) & 1.5     & scale factor by which \sex\ isophote is enlarged (for
neighbouring source treatment)\\
D05) & 20      & Definition of sky isophotes: additional offset to scale factor
(in pix), for sky measurement\\
D06) & 30      & Definition of sky isophotes: distance between individual sky
isophotes\\
D07) & 60      & Definition of sky isophotes: width of individual sky
isophotes\\
D08) & 30      & Definition of sky isophotes: gap between \sex\ isophote and
inner sky isophote\\
D09) & 3       & cut below which objects are considered as contributing to the
actual primary source\\
D10) & 2       & max number of allowed contributing sources per primary source\\
D11) & 1.4     & power by which the flux\_radius is raised to be converted to a
half-light radius\\
D12) & 5       & fraction of sources to be treated first (in \%; ``5''=5\%),
using multiple CPUs\\
D13) & 15      & calculate the slope of the sky from the x last determinations\\
D14) & -0.3    & slope in FWHM\_IMAGE vs. MAG\_BEST below which an object is
considered a star. Used for treating secondary sources\\
D15) & 6.8     & zeropoint in FWHM\_IMAGE vs. MAG\_BEST below which an object
is considered a star. Used for treating secondary sources\\
D16) & 5       & magnitude faint end limit for secondaries when fitting
galaxies. Objects more than x mag fainter than the primary are not included as
secondary sources but tertiaries\\
D17) & 2       & magnitude faint end limit for secondaries when fitting stars.
See also D16)\\
D18) & 8       & number of neighbouring tiles. See Sec.~\ref{sec_opt}\\
D19) & 4       & maximum number of parallel processes for fitting the
brightest sources, defined by D12)\\
D20) & 360     & minimum distance (in arcseconds) between all
simultaneously fitted sources for D19). If current source in fitting queue is
closer, fitting is delayed until other sources are done and the criterium is
fulfilled\\
\hline\hline
\multicolumn{3}{c}{{\sc Galfit} Setup}\\
\hline
E00) & /path/to/galfit-binary/galfit      & path and filename of \galfit\
executable\\
E01) & /path/to/survey/setup/batchlist.XX & path and filename of a batch
list, used for parallel processing of several tiles. For an example see
Sec.~\ref{sec_setup_batchlist}\\
E02) & obj                                & object file preposition. E.g.~for
E02) = ``obj'', a global file preposition ``im1.'' (from file list) and \sex\
detection number ``234'', the output filename would be: ``im1.obj234''\\
E03) & v\_gf                              & preposition for \galfit\ output
files. E.g.~for E03) = ``v\_gf'', a global file preposition ``im1.'' (from file
list) and \sex\ detection number ``234'', the output filename would be:
``im1.v\_gf234.fits''. Note, \galfit\ output files contain 3 FITS extensions:
the original image, the model (including fit parameters in the header) and the
residual image\\
E04) & /path/to/survey/setup/psf.fits     & PSF filename including path\\
E05) & mask                               & mask file preposition used in
\galfit. E.g.~for E05) = ``mask'', a global file preposition ``im1.'' (from
file list) and \sex\ detection number ``234'', the output filename would be:
``im1.mask234.fits''.\\
E06) & constr                             & constraint file preposition.
E.g.~for E06) = ``constr'', a global file preposition ``im1.'' (from file list)
and \sex\ detection number ``234'', the output filename would be:
``im1.constr234''.\\
E07) & 257                                & size of PSF convolution box\\
E08) & 26.486                             & magnitude zeropoint\\
E09) & 0.03                               & plate scale of the images
[arcsec/pixel]\\
E10) & 706                                & effective exposure time (after
image reduction, multidrizzling, etc.)\\
E11) & 750                                & constraint: maximum allowed
half-light radius\\
E12) & -5                                 & constraint: minimum magnitude
deviation (minus) from \sex\ measurement, \ie\ the fit magnitude is constrained
to not more than E12) mag brighter than the \sex\ value\\
E13) & 5                                  & constraint: maximum magnitude
deviation (plus) from \sex\ measurement. See also E12)\\
E14) & notnice                            & use the UNIX facility ``nice''
when starting the fitting with \galfit. Set E14) to ``nice'' to activate
``nicing''\\
E15) & 2.1c                               & \galfit\ version string.
E.g.~2.0.3c\\
\hline\hline
\multicolumn{3}{c}{Output Catalogue Setup}\\
\hline
F00) & execute      & execute catalogue combination block\\
F01) & combcat.fits & filename of combined FITS output catalogue. Output
directory is A01)\\
\hline
\end{supertabular}

%\clearpage
%\twocolumn
%new text in two-column mode...

%\bsp

\label{lastpage}

\end{document}